\documentclass[3p,times,twocolumn]{elsarticle}

\usepackage{ecrc}

\volume{00}

\firstpage{1}

\journalname{Nuclear Physics B Proceedings Supplement}

\runauth{}

\jid{nuphbp}

\jnltitlelogo{Nuclear Physics B Proceedings Supplement}

\usepackage{amssymb}

\usepackage[figuresright]{rotating}

\input babarsym.tex

\newcommand{\eett}   {\ensuremath{e^+e^- \to \tautau}\xspace}

\newcommand{\roots}        {\ensuremath{\sqrt{s}}\xspace}

\newcommand{\tautohhh} {\ensuremath{\tau^- \to h^-h^-h^+\nu_\tau}\xspace}
\newcommand{\tautoppp} {\ensuremath{\tau^- \to \pi^-\pi^-\pi^+\nu_\tau}\xspace}
\newcommand{\tautokpp} {\ensuremath{\tau^- \to K^-\pi^-\pi^+\nu_\tau}\xspace}
\newcommand{\tautokpk} {\ensuremath{\tau^- \to K^-\pi^-K^+\nu_\tau}\xspace}
\newcommand{\tautokkk} {\ensuremath{\tau^- \to K^-K^-K^+\nu_\tau}\xspace}

\newcommand{\tautophik} {\ensuremath{\tau^- \to \phik   \nu_\tau}\xspace}

\newcommand{\phik} {\ensuremath{\phi K^-}\xspace}

\newcommand{\tautokpppz} {\ensuremath{\tau^- \to K^-\pi^-\pi^+\pi^{0}\nu_\tau}\xspace}
\newcommand{\tautokpkpz} {\ensuremath{\tau^- \to K^-\pi^-K^+\pi^{0}\nu_\tau}\xspace}

\newcommand{\lumi}             {\ensuremath{342\invfb}\xspace}

\def\kk2f       {\mbox{\tt KK}\xspace}
\def\tauola     {\mbox{\tt Tauola}\xspace}

\begin{document}

\begin{frontmatter}



\dochead{}

\title{Invariant mass spectra of \tautohhh decays}


\author{I. M. Nugent[on behalf of the \babar\ Collaboration] }

\address{III. Physikalisches Institut, Physikzentrum, RWTH Aachen,
  52056 Aachen}
\address{Department of Physics and Astronomy, University of Victoria,
PO Box 3055, STN CSC, Victoria, BC, V8W 3P6 Canada}

\begin{abstract}
Using \lumi of \epem annihilation data collected with the \babar\
detector at the SLAC PEP-II electron-positron asymmetric energy
collider operating at a
center-of-mass energies near 10.58\gev, we present the preliminary
measurements of the invariant mass
distributions of
$\tautoppp $,
$\tautokpp $,
$\tautokpk $ and
$\tautokkk $, where events with $K^0_S\rightarrow \pi^+\pi^-$ decays
are excluded.

\end{abstract}

\begin{keyword}
Tau, Invariant-Mass 
\end{keyword}

\end{frontmatter}


\section{Introduction}
\label{Introduction}

Hadronic $\tau$ decays provide an opportunity to measure the coupling
strength of the weak current to the first
and second generations of quarks~\cite{Cabibbo:1963yz}. They also
provide a
clean environment to probe low energy QCD and measure fundamental
properties of the Standard Model of Particle Physics. Hadronic $\tau$
decays also
play a critical role as a probe to search for new physics at the
LHC. With
the recent data collected at the B-Factory detectors, significant
improvements in our knowledge of the hadronic decay structure can be
made, allowing for improved modeling of $\tau$ decays at the LHC
 and at the next generation B-Factories.

\section{Detector, Data Sample and Monte Carlo samples}
\label{DetectorDataMC}
This analysis employs data collected with the \babar\ detector at the
PEP-II storage ring which
has a centre-of-mass (CM) energy near (\roots) near 10.58\gev. At these
energies, the cross section is
$\sigma_{\eett}=(0.919\pm0.003)$nb~\cite{kk}. A detailed description of the \babar\
detector and MC Simulation
can be found in ~\cite{detector} and ~\cite{ourpaper}.
 

\section{Invariant Mass Measurements}
\label{InvariantMass}

The analysis of the invariant mass spectra presented in this work is a
continuation of branching fraction measurements from \cite{ourpaper}.
Therefore, the work presented here uses the event
selection procedure from \cite{ourpaper}, but incorporates results
from additional studies. For the event selection, a sample of
\tautohhh decays from
 \eett events is selected by requiring the partner $\tau^+$ to decay
 leptonically.
 Within this sample,
each of the $h^{\pm}$ mesons is uniquely identified as a charged pion
 or kaon, and the decay
categorized as \tautoppp, \tautokpp, \tautokpk or \tautokkk, where
 events with
$K^0_S \ra \pi^+\pi^-$ have been excluded.
An efficiency correction, initially
 obtained from MC and corrected using data control samples,  is used
to correct for efficiency losses from the event selection
 for each interval in the invariant mass distribution.
A detailed description of the selection can be found in
\cite{ourpaper}.
 After events are selected
the invariant mass distributions are analyzed. First, an arithmetic
subtraction of the backgrounds is applied to the
invariant mass distributions for
 each channel. The \tautohhh backgrounds between the channels caused
 by  particle mis-identification (cross-feed) are normalized to the
 measured branching fractions in \babar\
 \cite{ourpaper}.
 Detector effects are then removed using Bayesian Unfolding
 \cite{D'Agostini:1994zf}, which has been trained using the signal MC
 for each decay mode. The invariant
 mass distributions are then normalized to unity.

\section{Results}
\label{Results}

\begin{figure*}
\begin{center}
\resizebox{300pt}{99pt}{
\includegraphics[bb = 0pt 0pt 570pt 535pt]{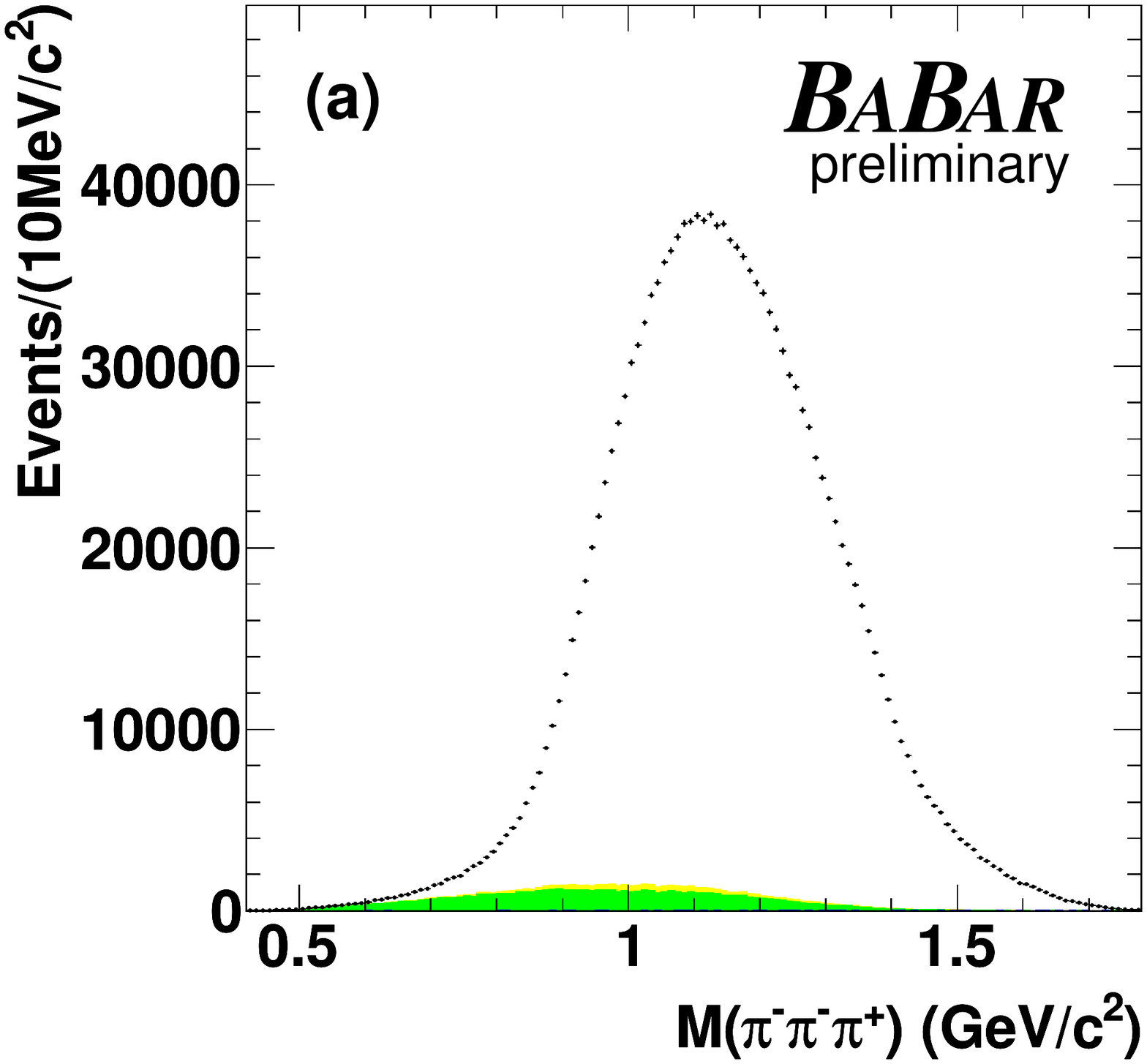}
\includegraphics[bb = 0pt 0pt 570pt 535pt]{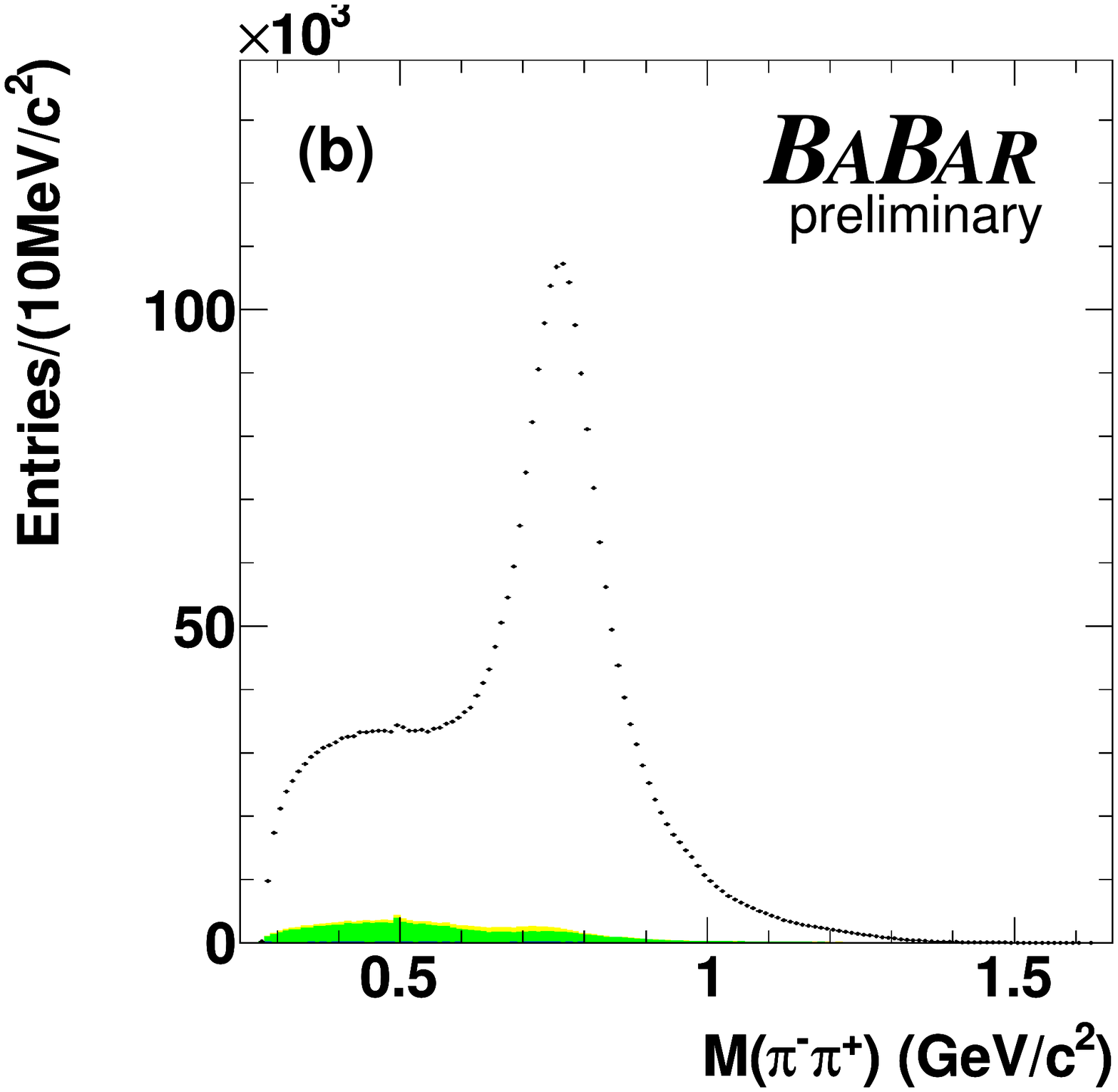}
\includegraphics[bb = 0pt 0pt 570pt 535pt]{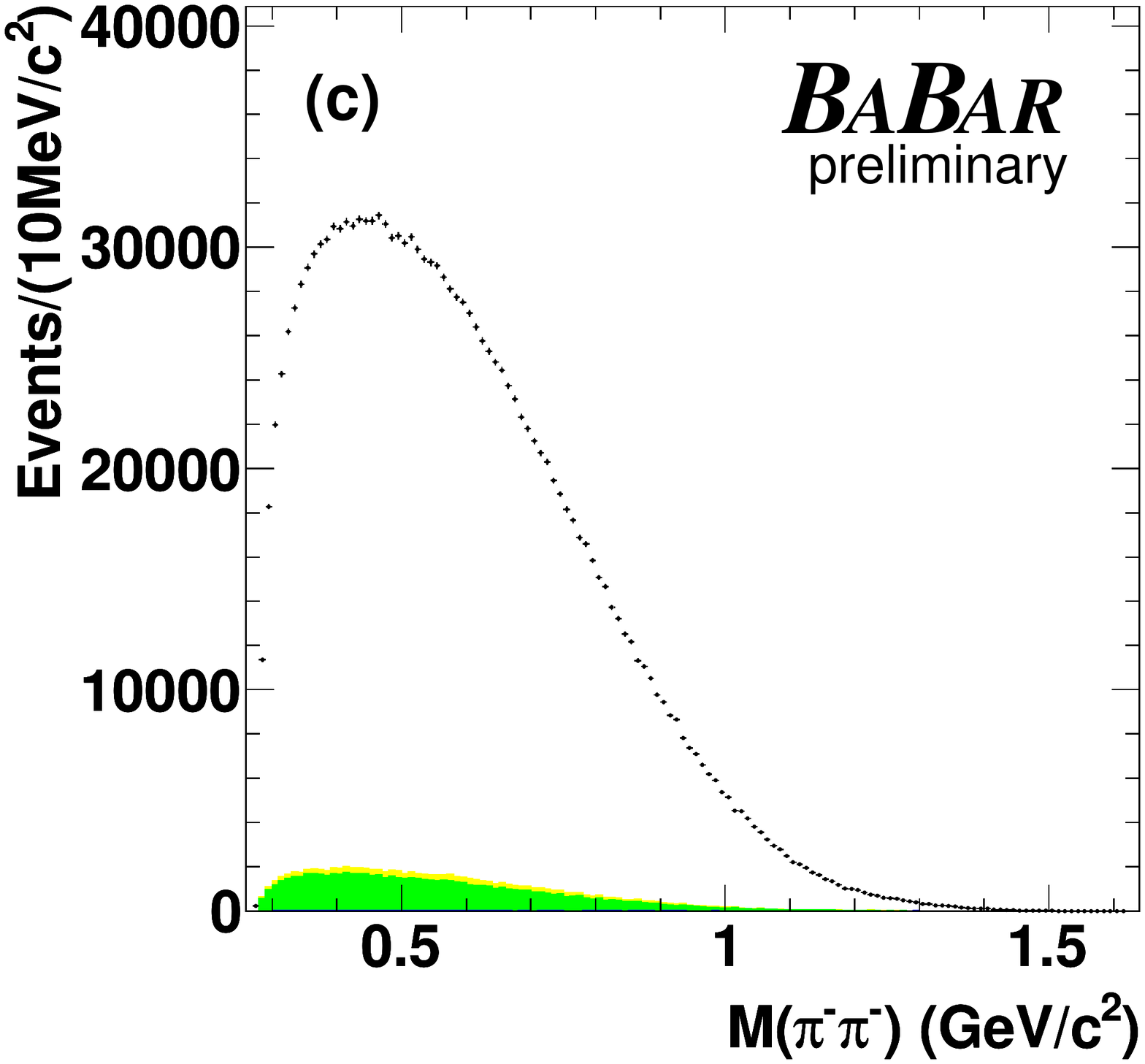}
}
\resizebox{300pt}{99pt}{
\includegraphics[bb = 0pt 0pt 570pt 535pt]{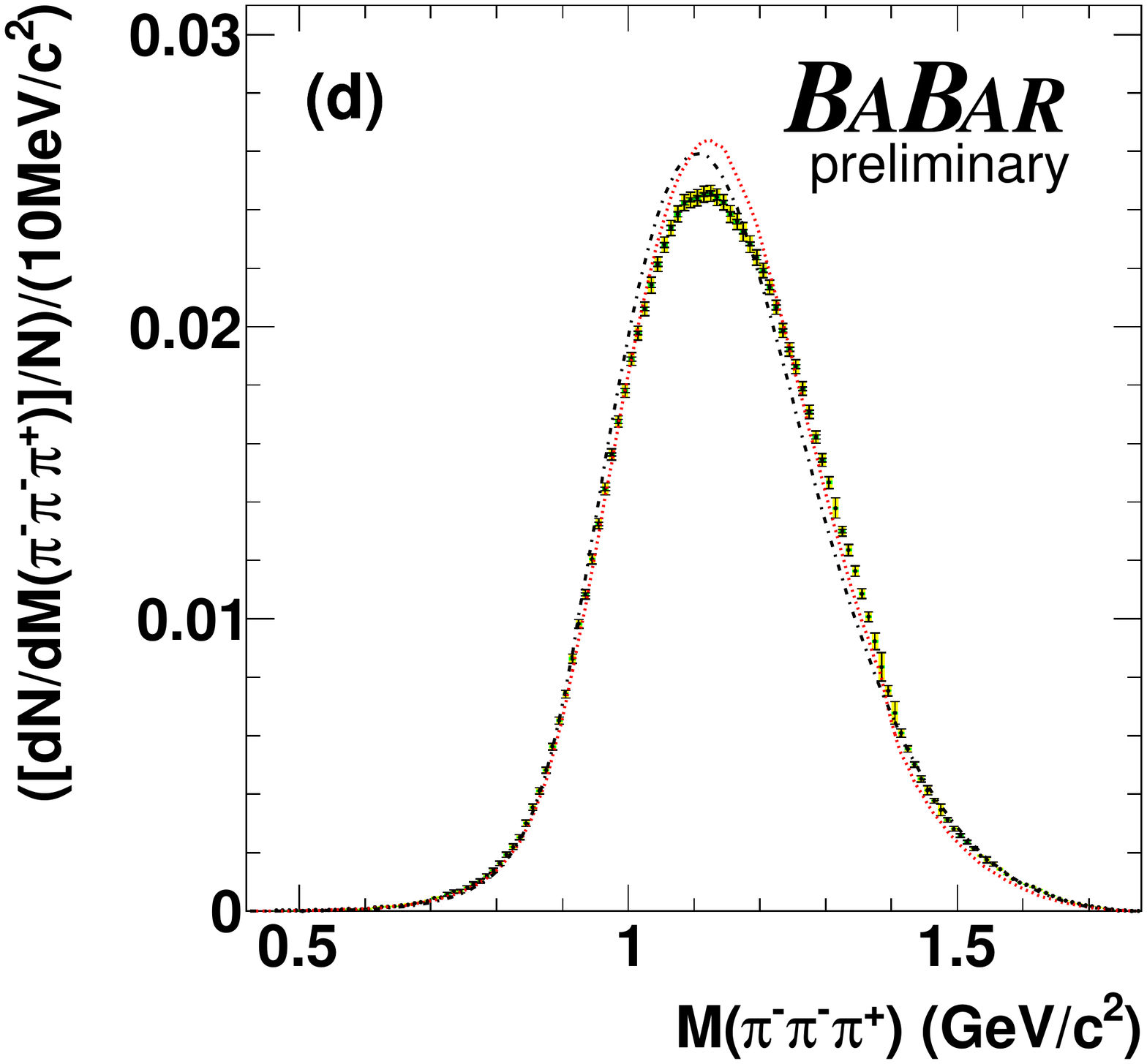}
\includegraphics[bb = 0pt 0pt 570pt 535pt]{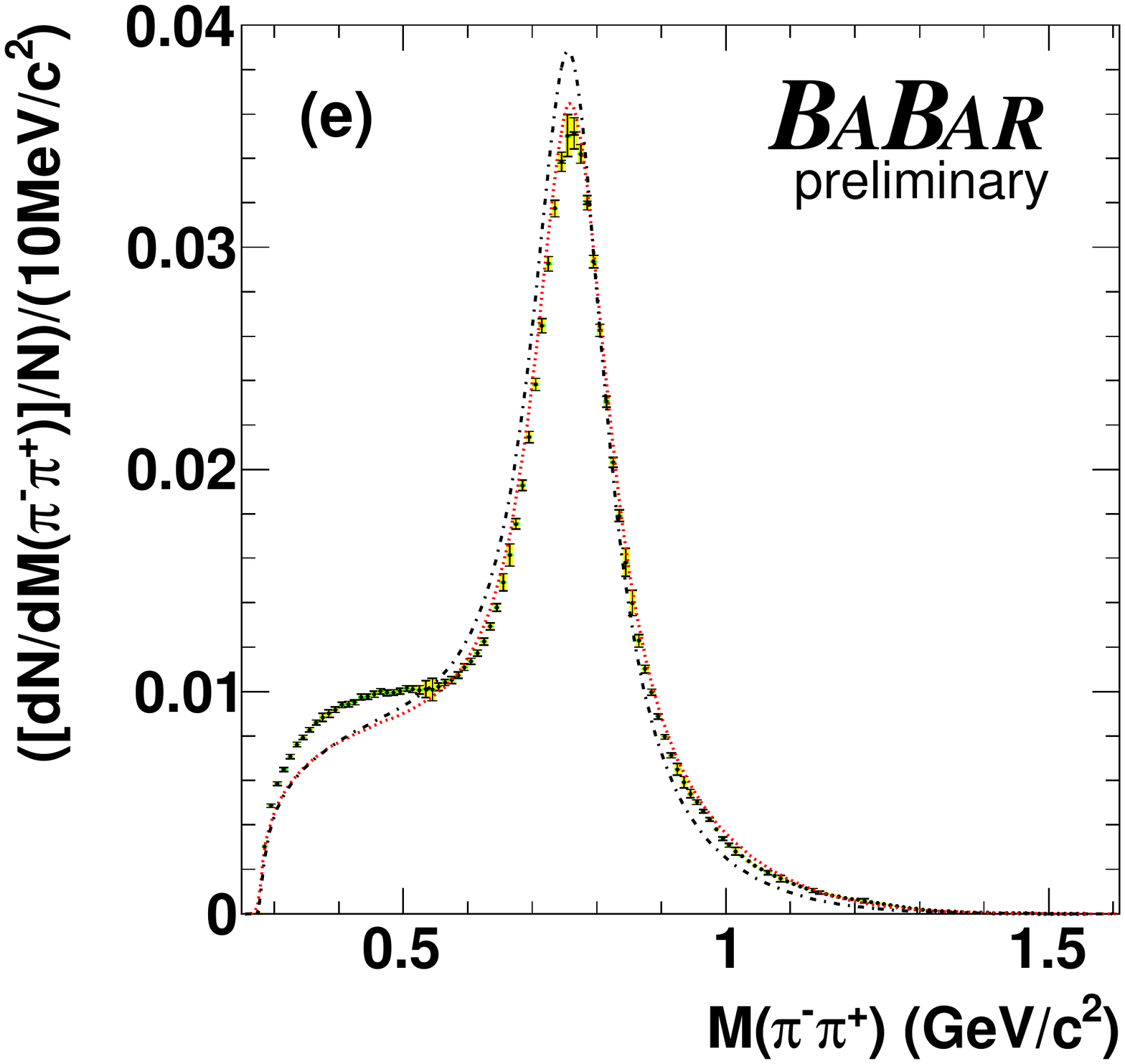}
\includegraphics[bb = 0pt 0pt 570pt 535pt]{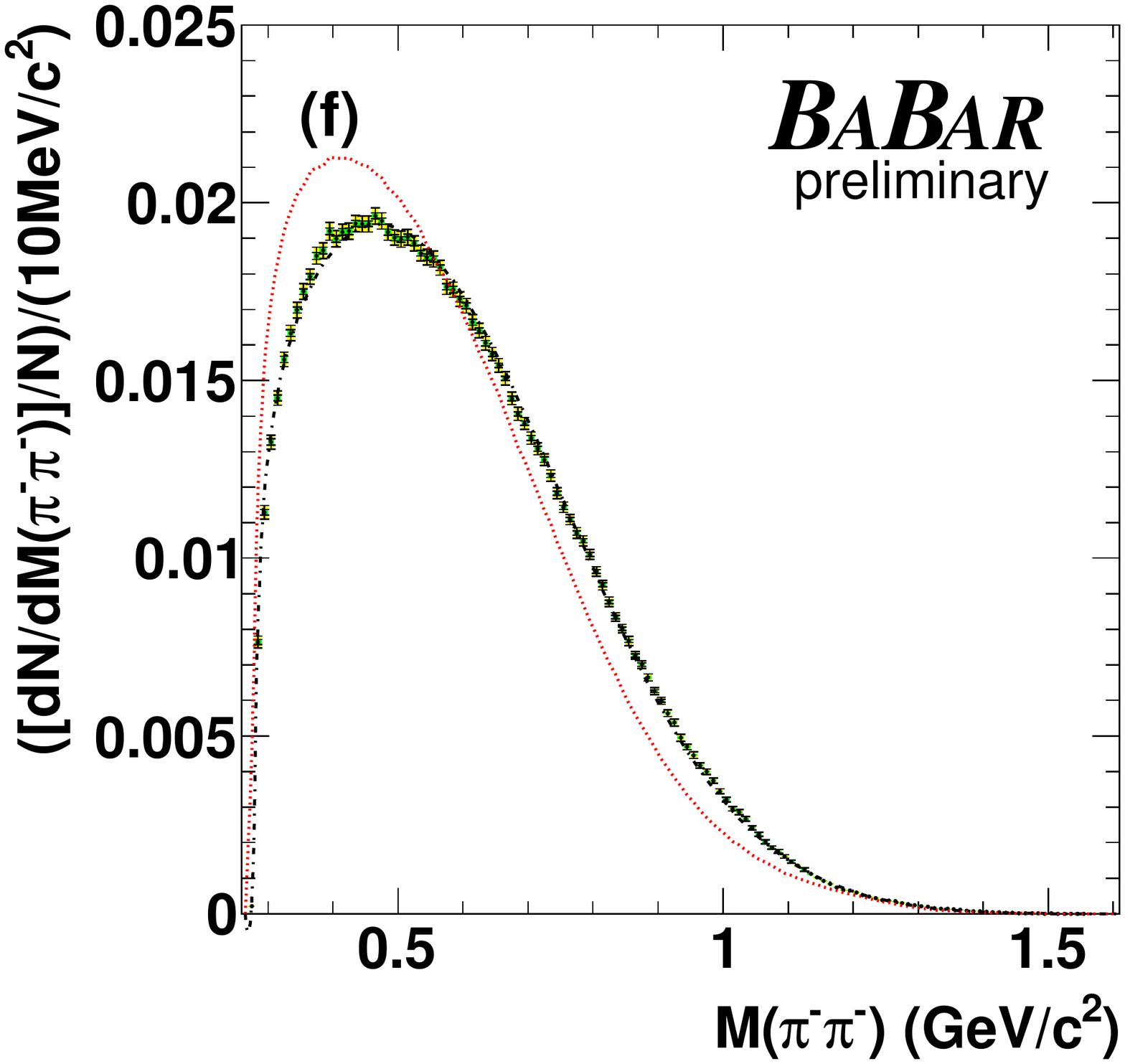}
}
\caption{ The reconstructed and unfolded invariant mass spectra for
the \tautoppp channels.  The reconstructed invariant mass
distributions for (a)
$M(\pi^{-}\pi^{-}\pi^{+})$, (b)  $M(\pi^{-}\pi^{+})$ and (c)
$M(\pi^{-}\pi^{-})$ are
presented in the first row. In the second row, the (d)
$M(\pi^{-}\pi^{-}\pi^{+})$, (e)  $M(\pi^{-}\pi^{+})$ and (f)
$M(\pi^{-}\pi^{-})$ unfolded invariant mass spectra are shown.
For the reconstructed mass plots, the data is represented by the
points with the error bars representing the statistical
uncertainty. The blue (dark)  histogram represents the non-$\tau$
background MC, the green (medium dark) histogram represents the $\tau$
backgrounds excluding the \tautohhh cross-feed  which
are represented by the yellow (light)  histogram. For the unfolded
mass plots,
the  data is represented by the
points with the inner error bars (green) representing the statistical
uncertainty and the outer error bars (yellow)
representing the statistical and systematic uncertainties added in
quadrature. The integral of the unfolded distribution has been
normalized to 1. The black dashed line is the generator level MC
distribution used in the \babar\ simulation. The red dotted line is
the
CLEO tune for \tauola 2.8~\cite{Golonka:2003xt}.}
\label{figure1_3pi}
\end{center}
\end{figure*}

\begin{figure*}
\begin{center}
\resizebox{400pt}{99pt}{
\includegraphics[bb = 0pt 0pt 570pt 535pt]{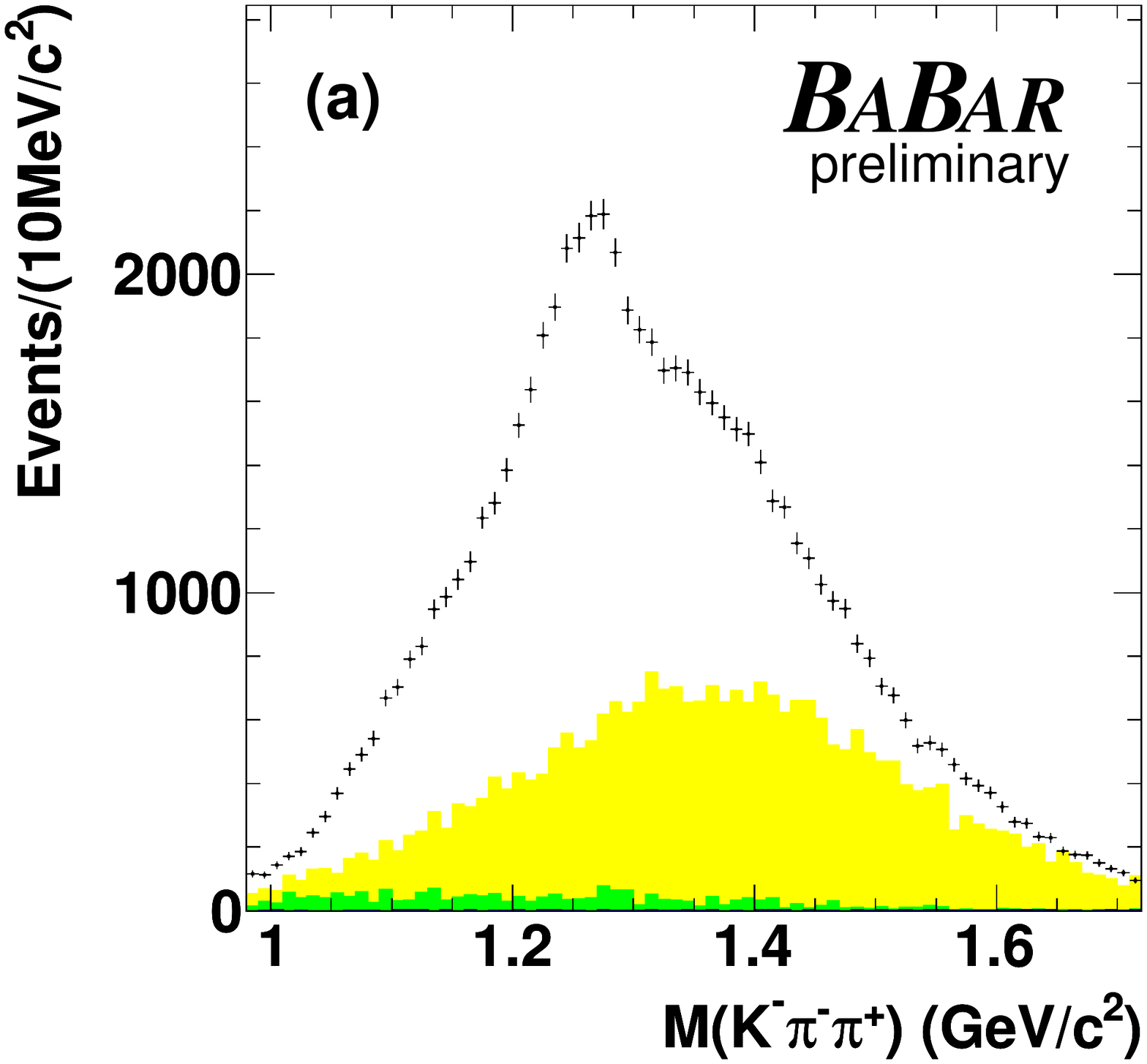}
\includegraphics[bb = 0pt 0pt 570pt 535pt]{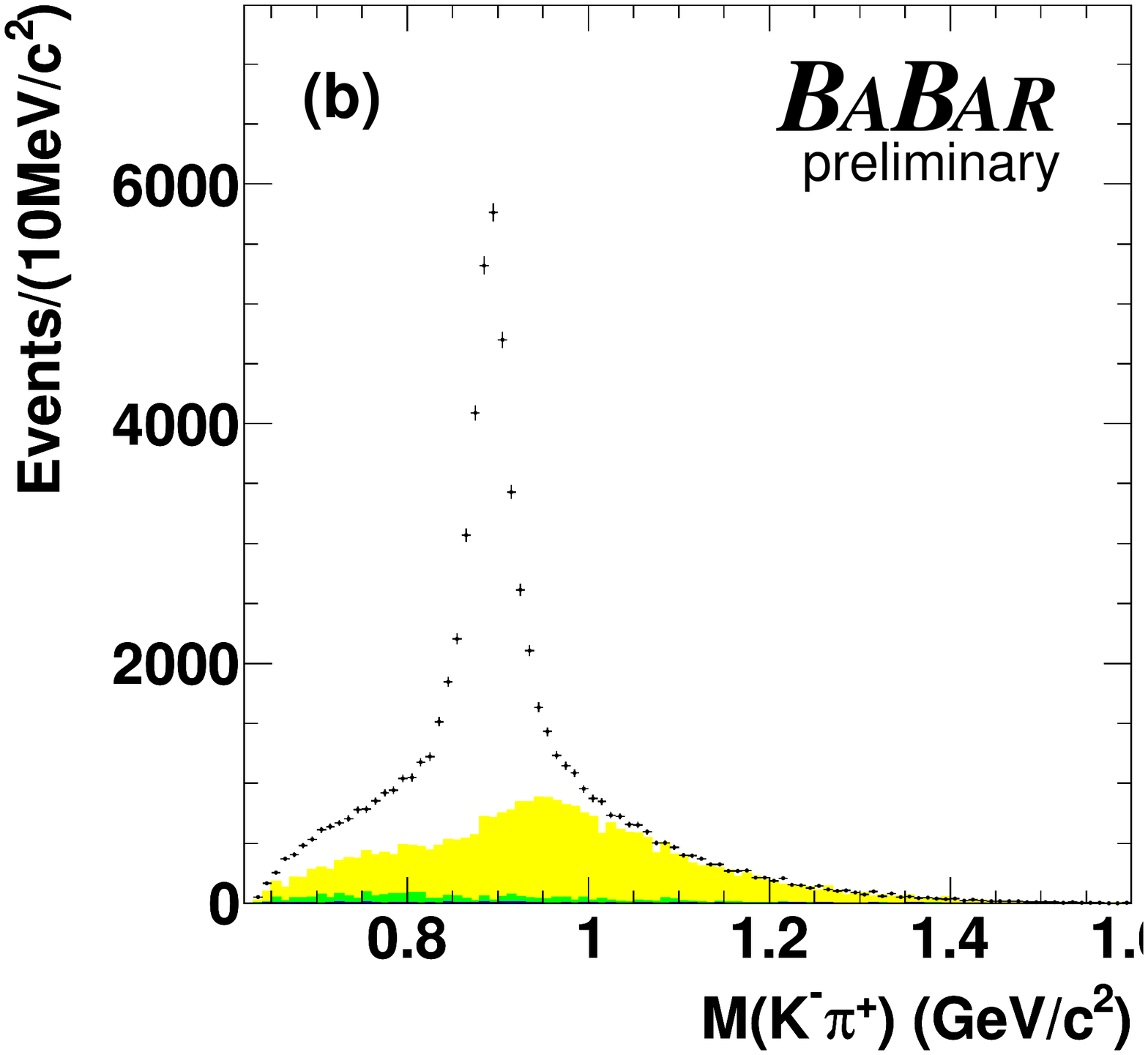}
\includegraphics[bb = 0pt 0pt 570pt 535pt]{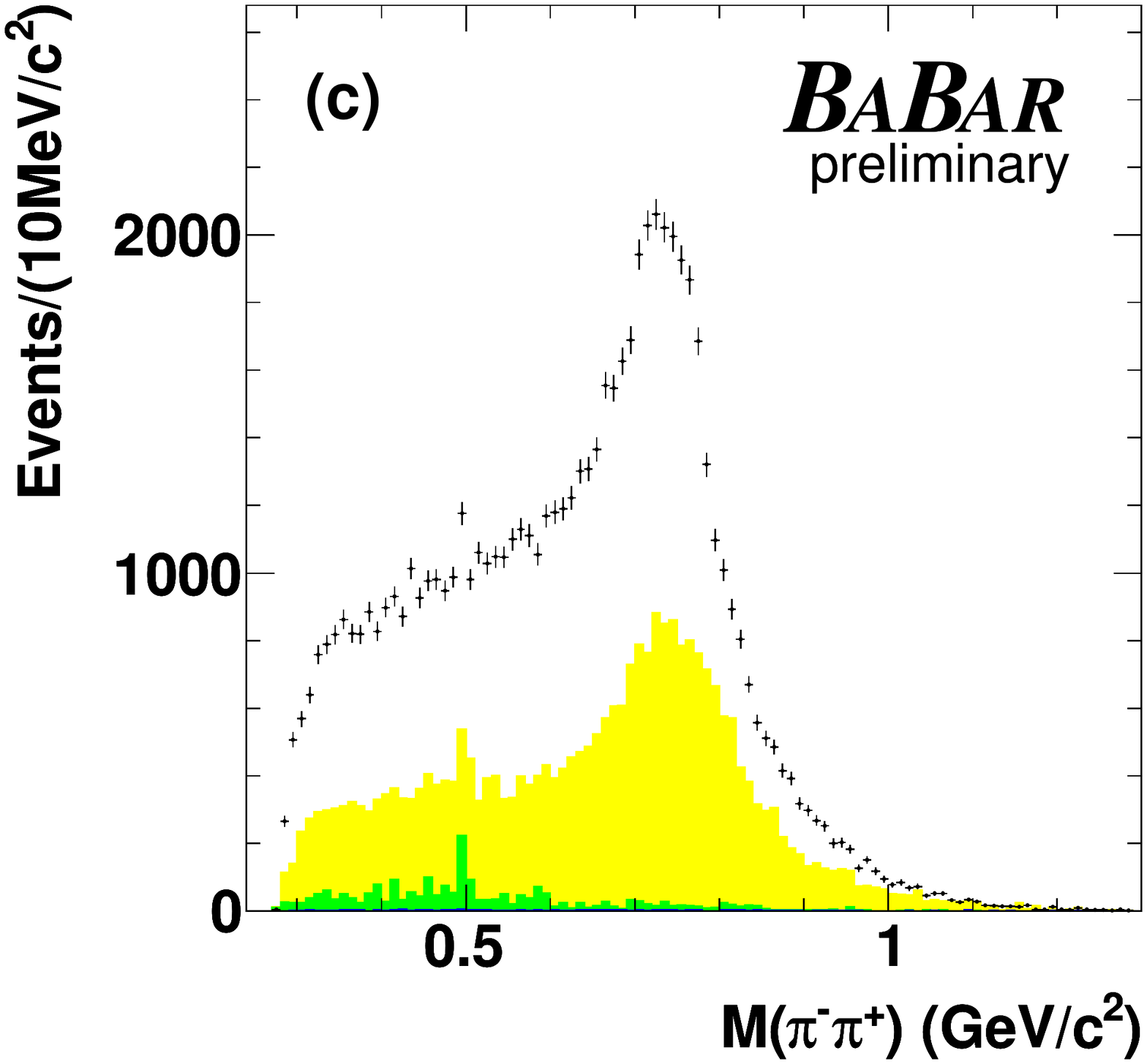}
\includegraphics[bb = 0pt 0pt 570pt 535pt]{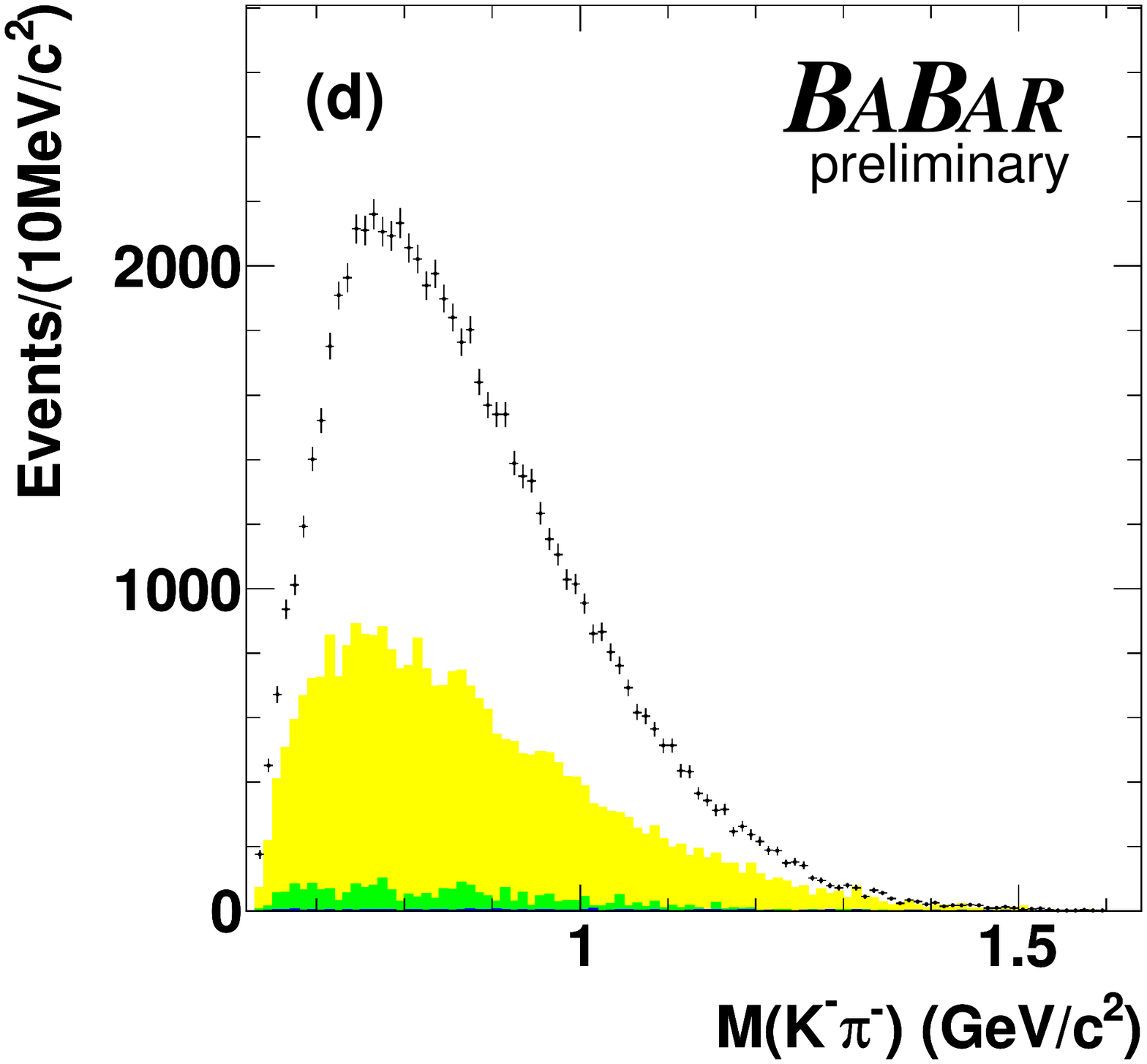}
}
\resizebox{400pt}{99pt}{
\includegraphics[bb = 0pt 0pt 570pt 535pt]{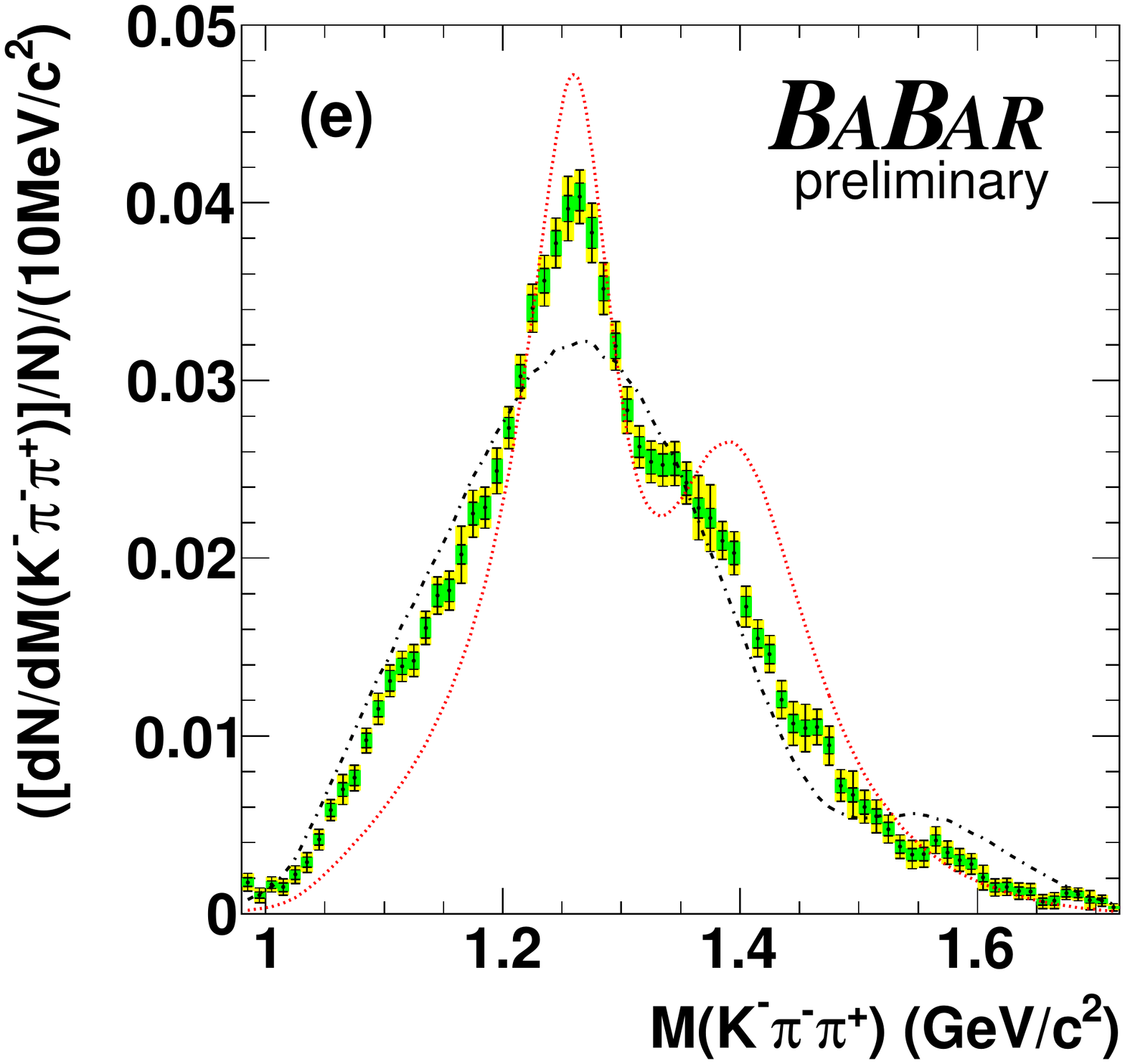}
\includegraphics[bb = 0pt 0pt 570pt 535pt]{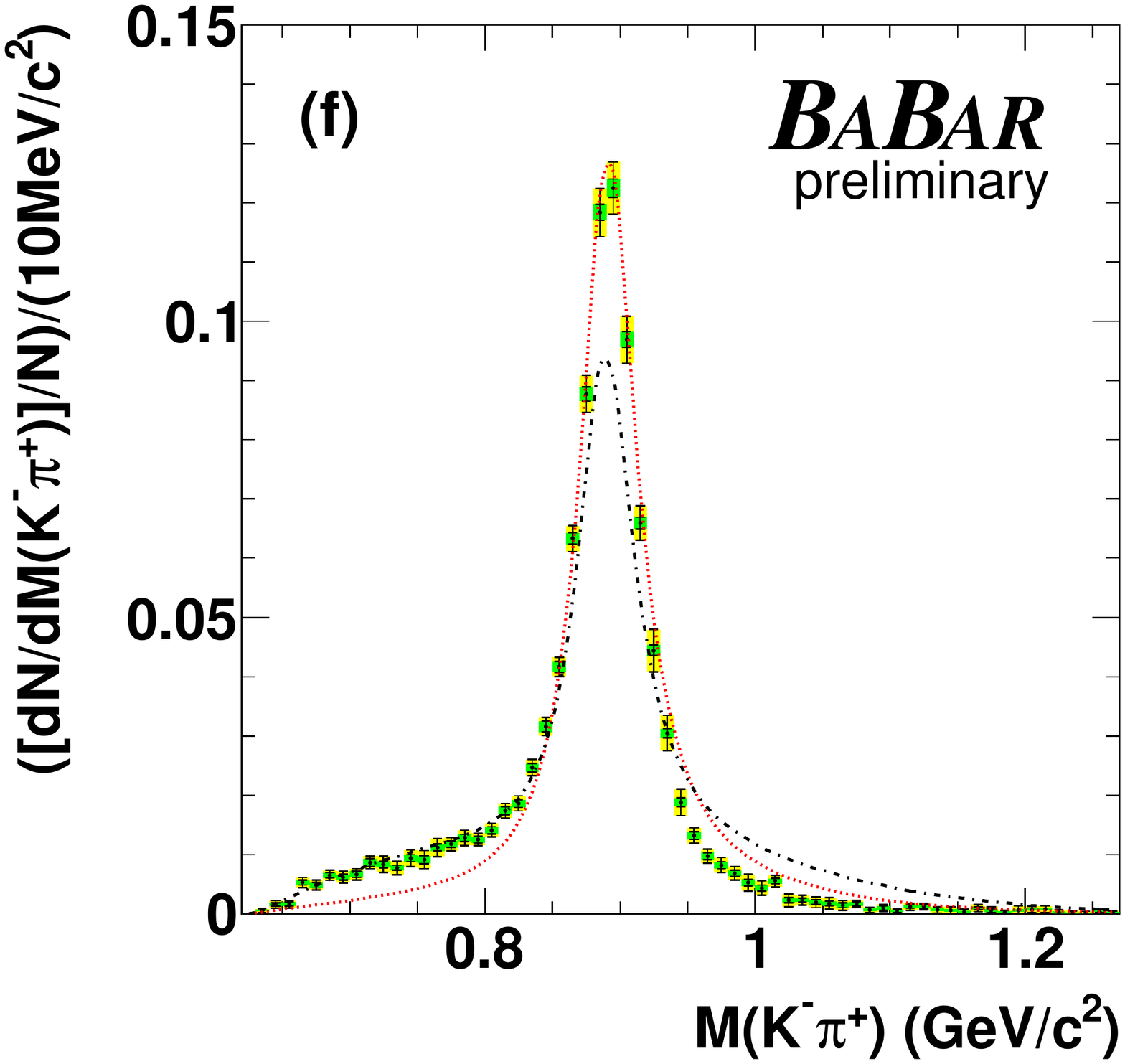}
\includegraphics[bb = 0pt 0pt 570pt 535pt]{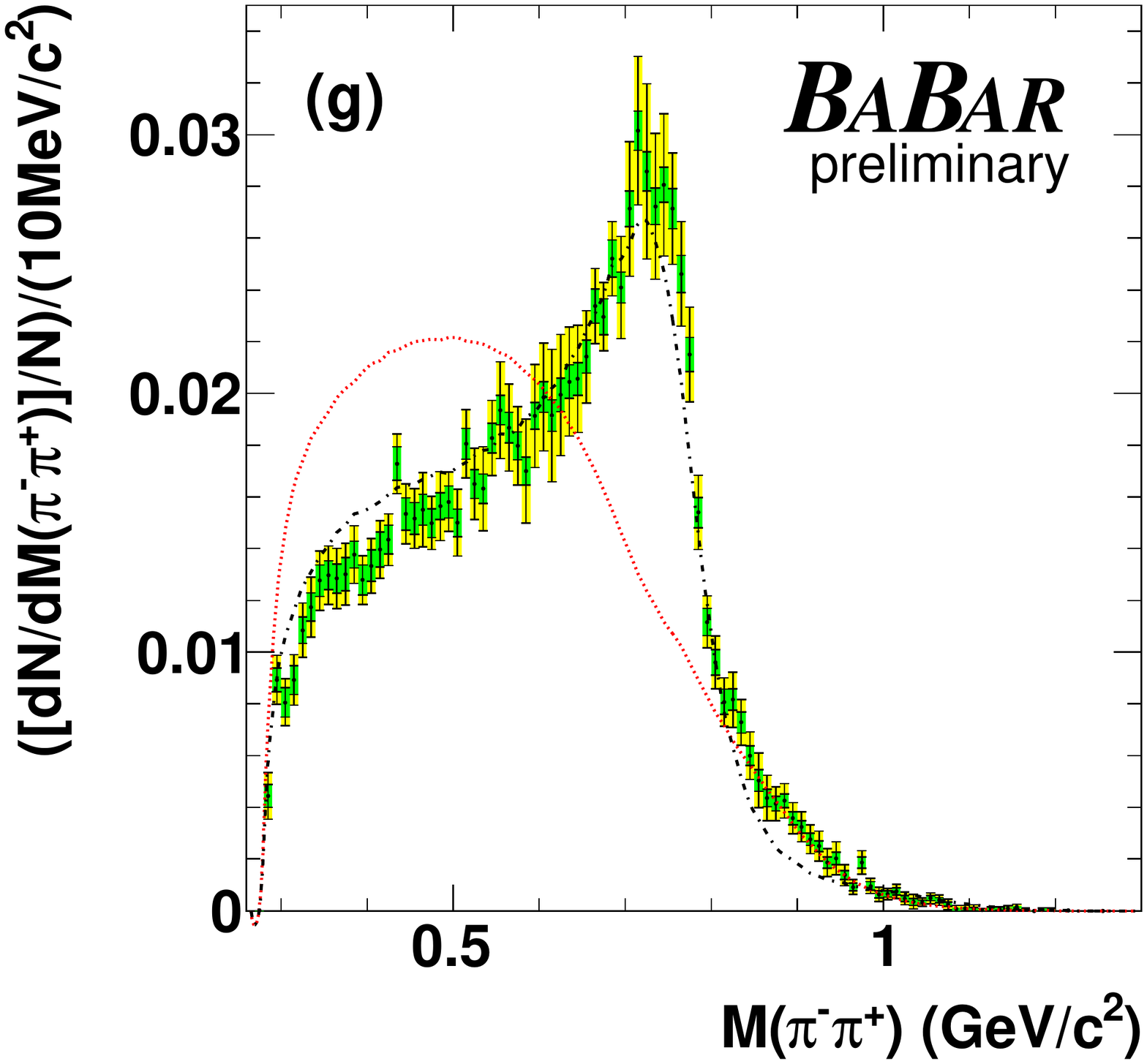}
\includegraphics[bb = 0pt 0pt 570pt 535pt]{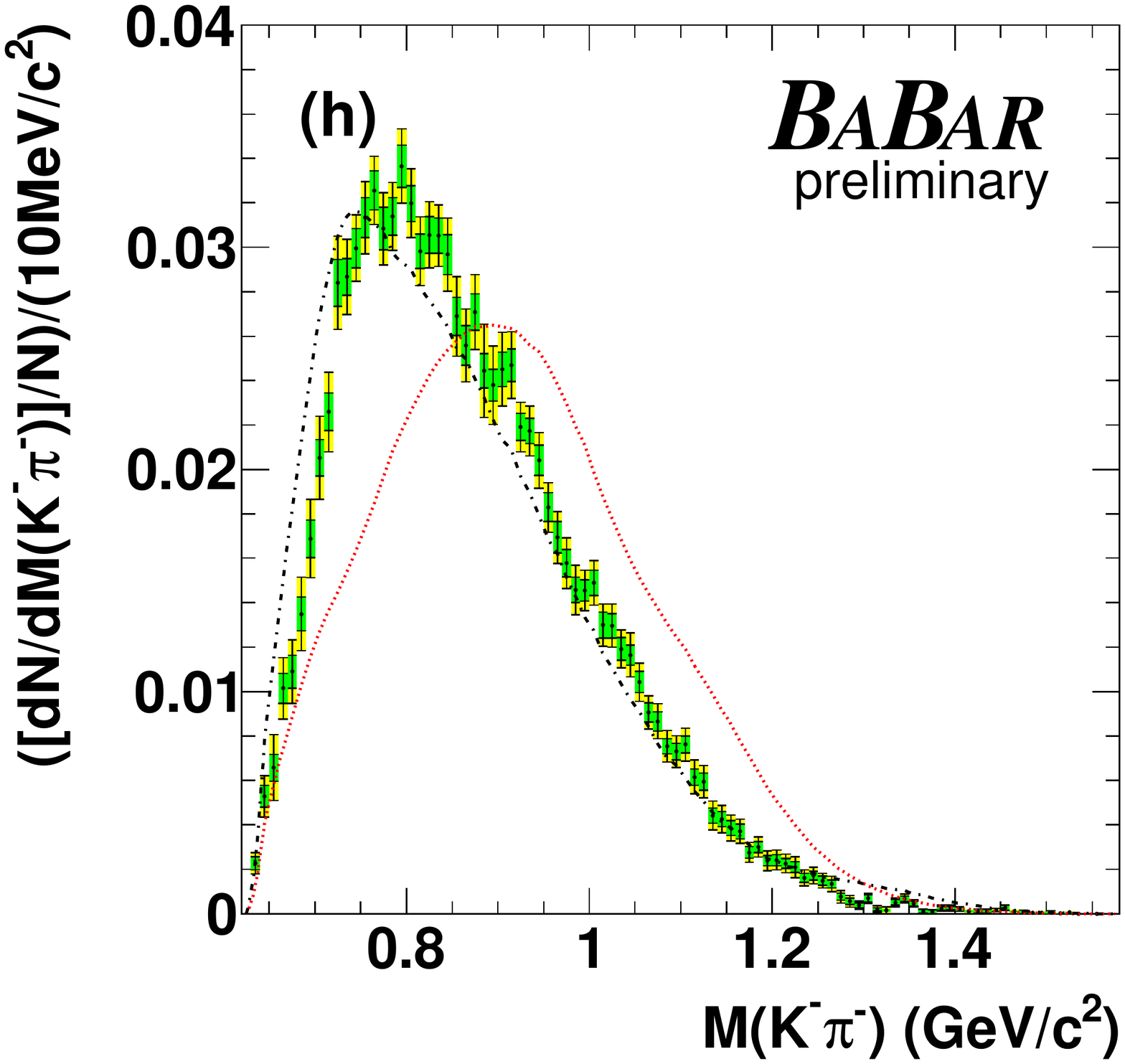}
}
\caption{ The reconstructed and unfolded invariant mass spectra for
the \tautokpp channels.  The reconstructed invariant mass
distributions for (a)
$M(K^{-}\pi^{-}\pi^{+})$, (b)  $M(K^{-}\pi^{+})$, (c)
$M(\pi^{-}\pi^{+ })$ and (d) $M(K^{-}\pi^{-})$ are
presented in the first row. In the second row, the (e)
$M(K^{-}\pi^{-}\pi^{+})$, (f)  $M(K^{-}\pi^{+})$, (g)
$M(\pi^{-}\pi^{+})$ and (h) $M(K^{-}\pi^{-})$  unfolded invariant mass
spectra are shown.
For the reconstructed mass plots, the data is represented by the
points with the error bars representing the statistical
uncertainty. The blue (dark)  histogram represents the non-$\tau$
background MC, the green (medium dark) histogram represents the $\tau$
backgrounds excluding the \tautohhh cross-feed  which
are represented by the yellow (light)  histogram. For the unfolded
mass plots,
the  data is represented by the
points with the inner error bars (green) representing the statistical
uncertainty and the outer error bars (yellow)
representing the statistical and systematic uncertainties added in
quadrature. The integral of the unfolded distribution has been
normalized to 1. The black dashed line is the generator level MC
distribution used in the \babar\ simulation. The red dotted line is
the
CLEO tune for \tauola 2.8~\cite{Golonka:2003xt}.}
\label{figure1_kpipi}
\end{center}
\end{figure*}

\begin{figure*}
\begin{center}
\resizebox{400pt}{99pt}{
\includegraphics[bb = 0pt 0pt 570pt 535pt]{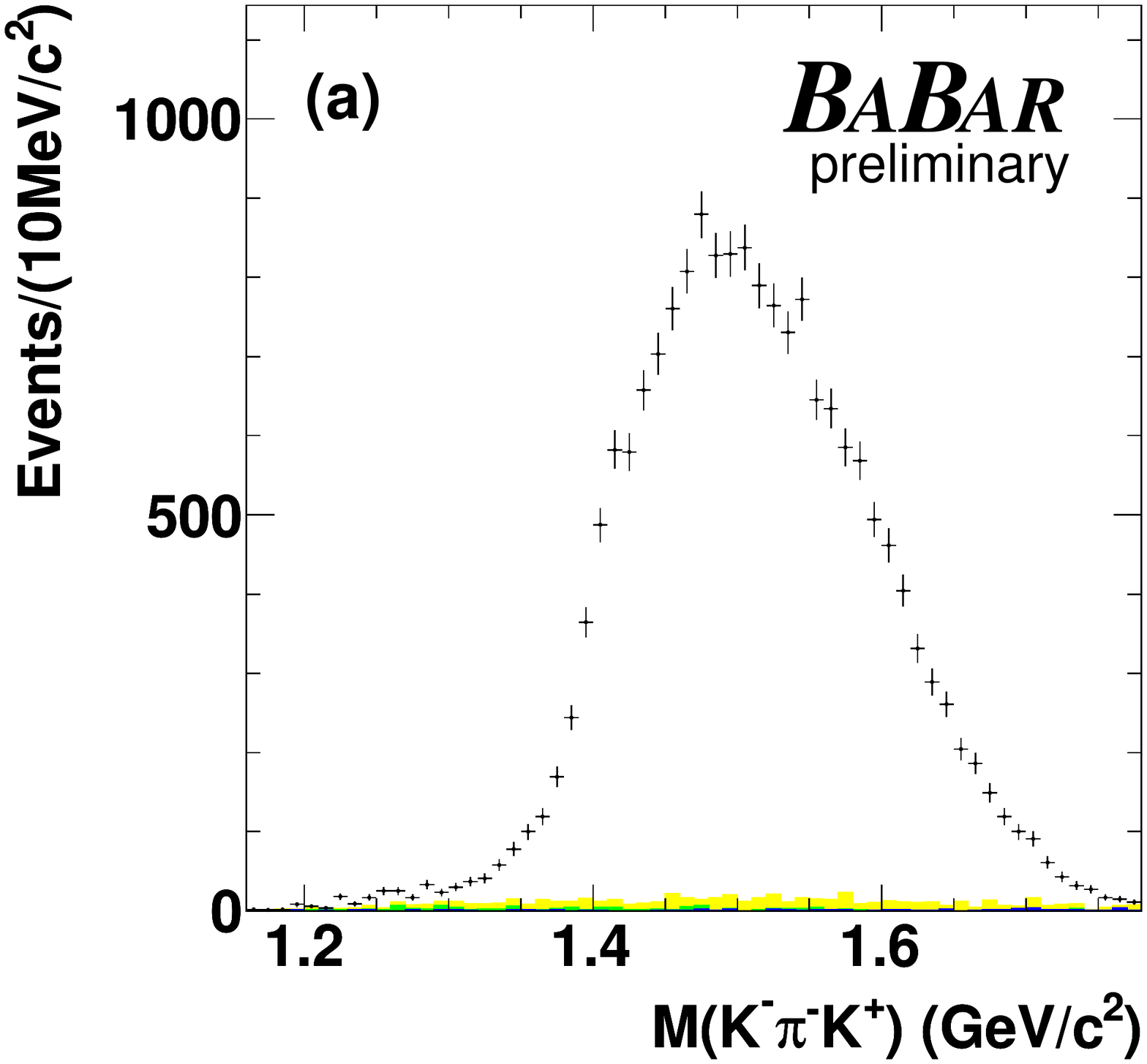}
\includegraphics[bb = 0pt 0pt 570pt 535pt]{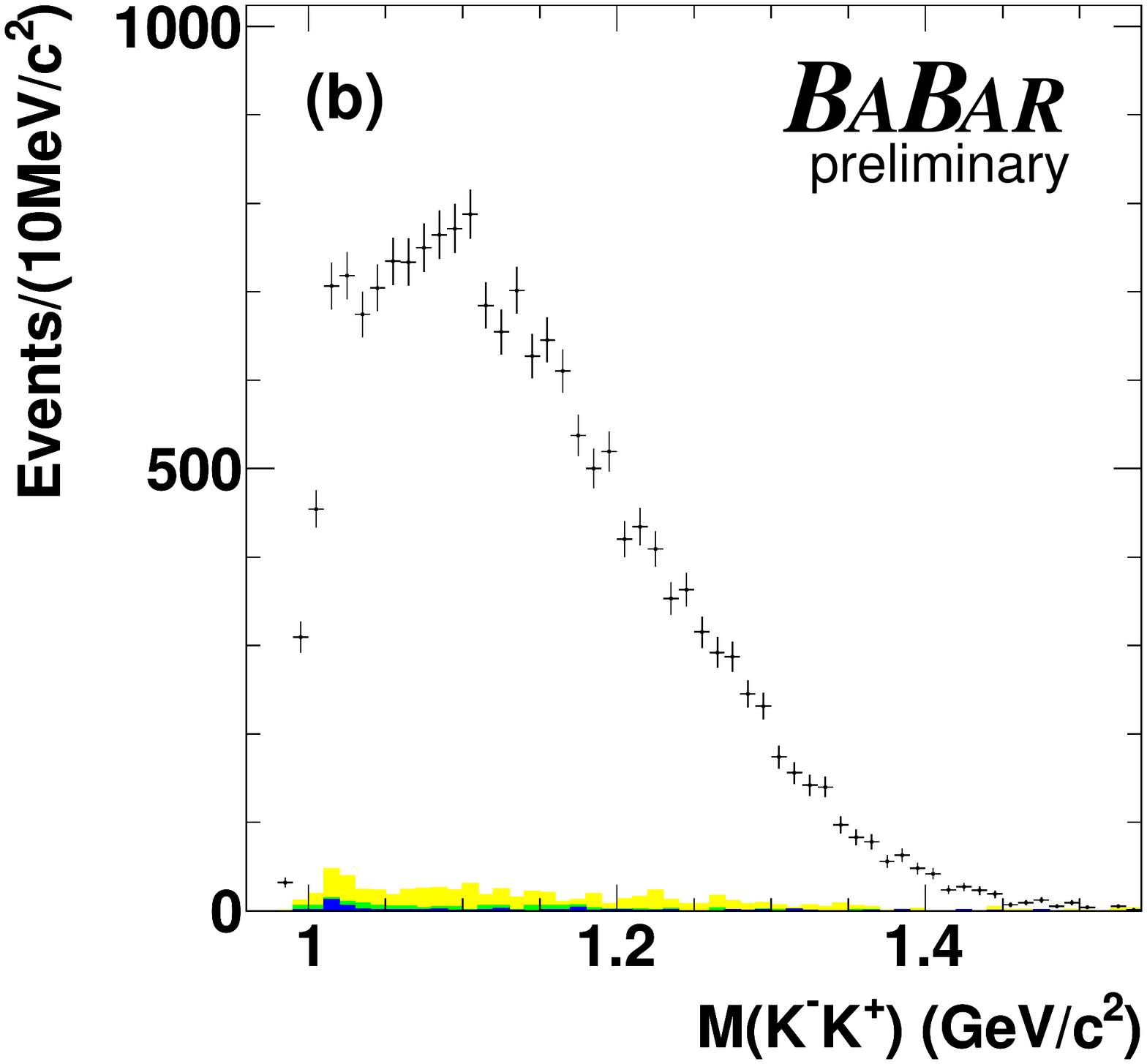}
\includegraphics[bb = 0pt 0pt 570pt 535pt]{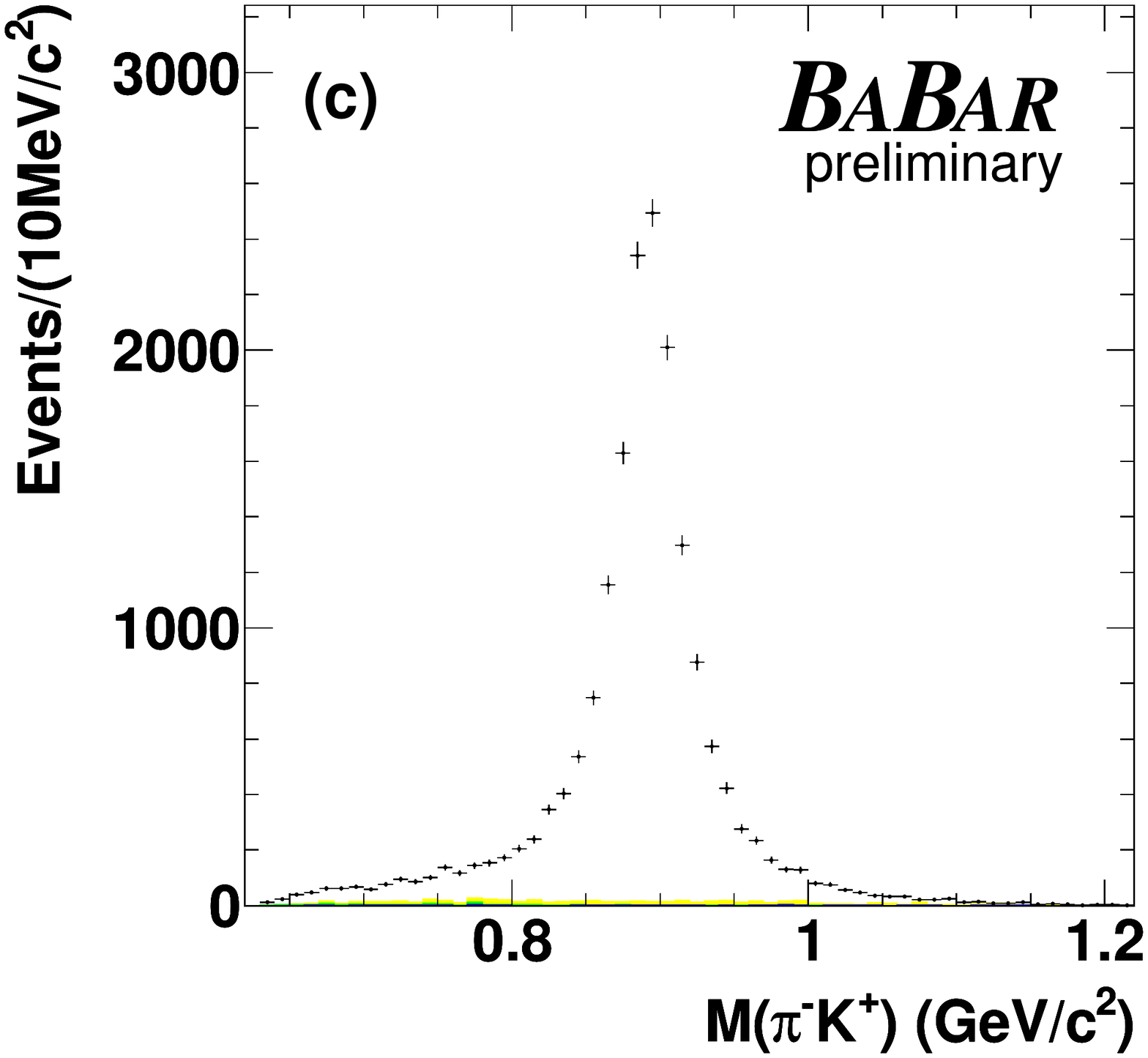}
\includegraphics[bb = 0pt 0pt 570pt 535pt]{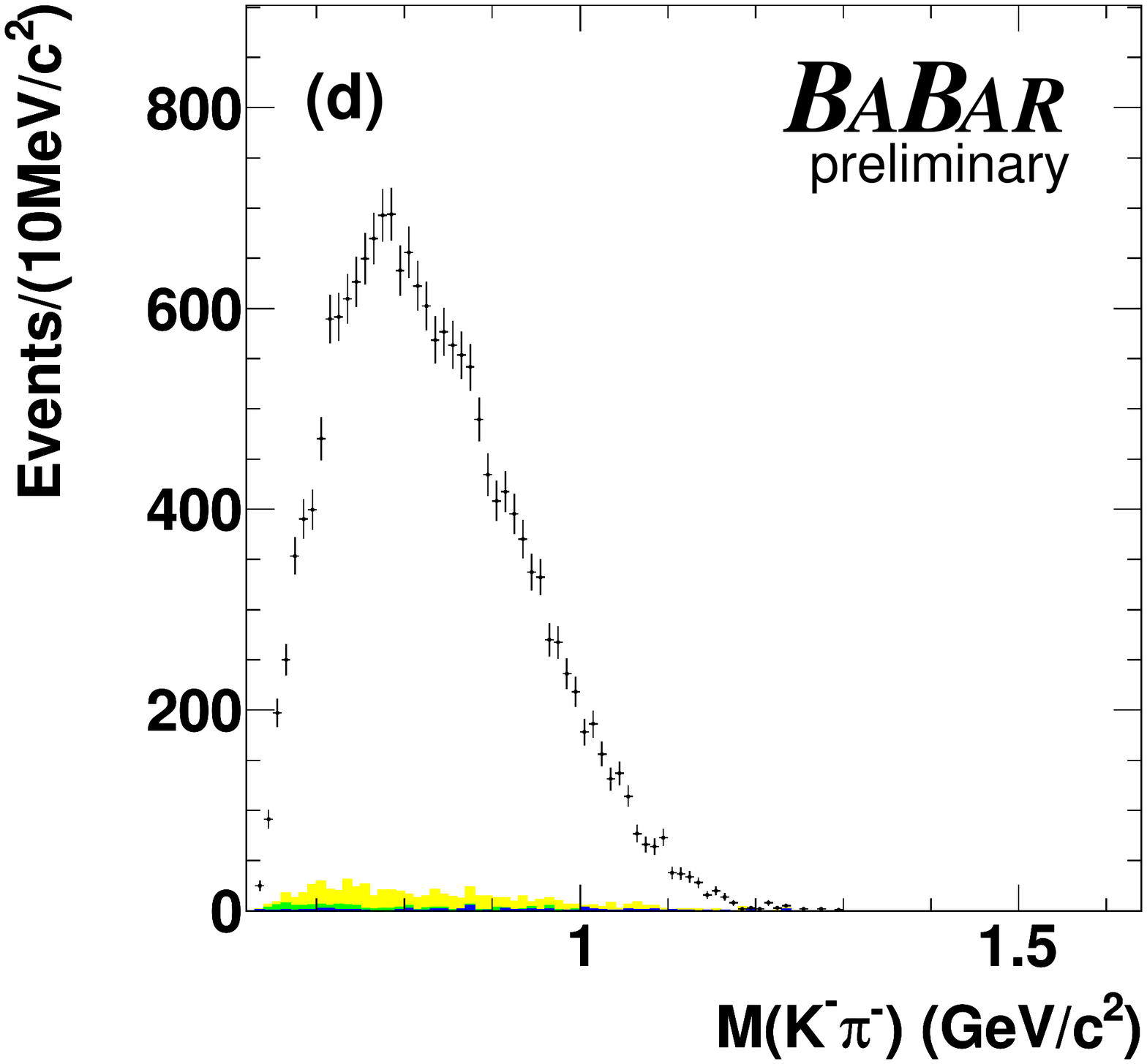}
}
\resizebox{400pt}{99pt}{
\includegraphics[bb = 0pt 0pt 570pt 535pt]{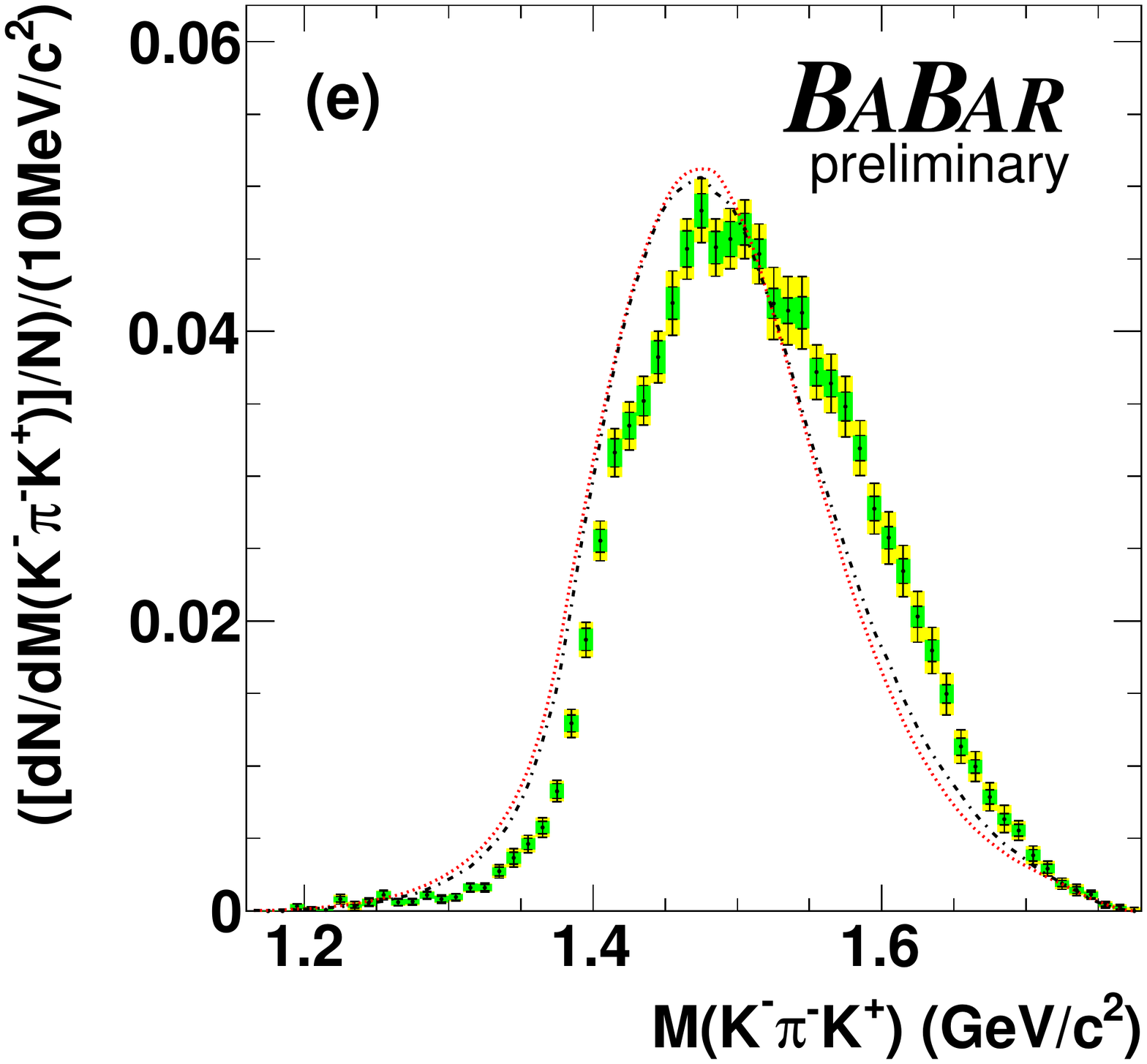}
\includegraphics[bb = 0pt 0pt 570pt 535pt]{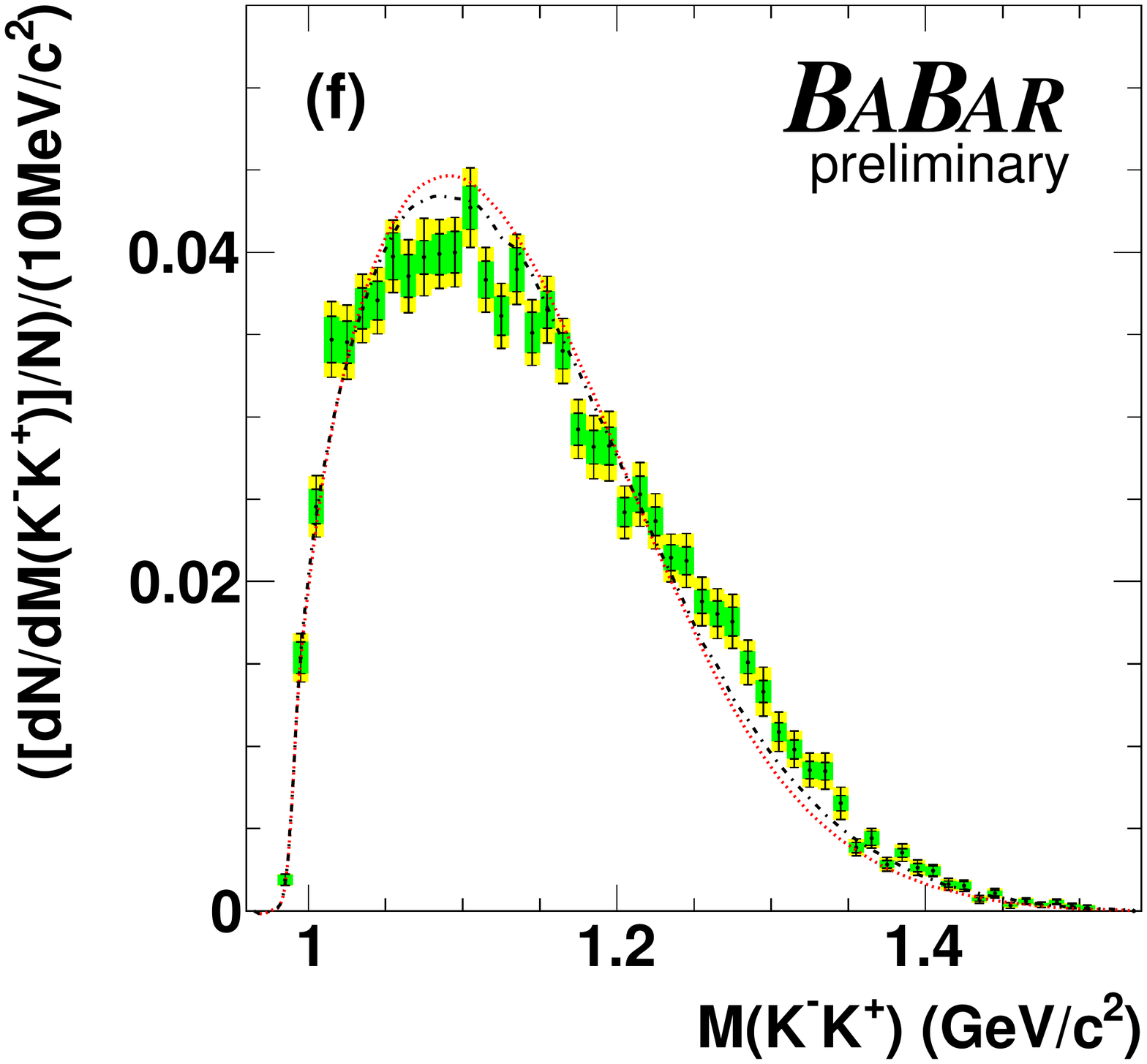}
\includegraphics[bb = 0pt 0pt 570pt 535pt]{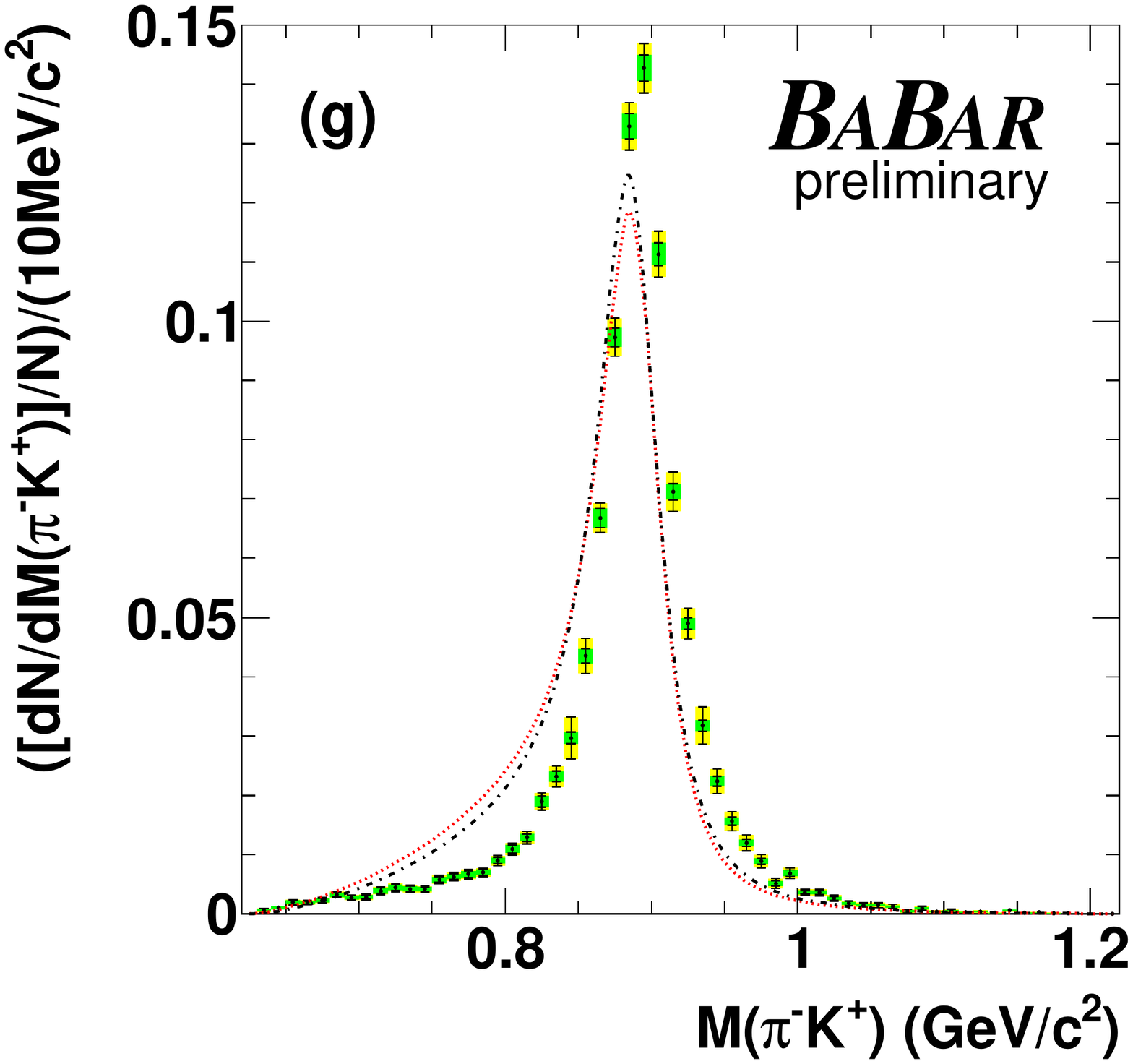}
\includegraphics[bb = 0pt 0pt 570pt 535pt]{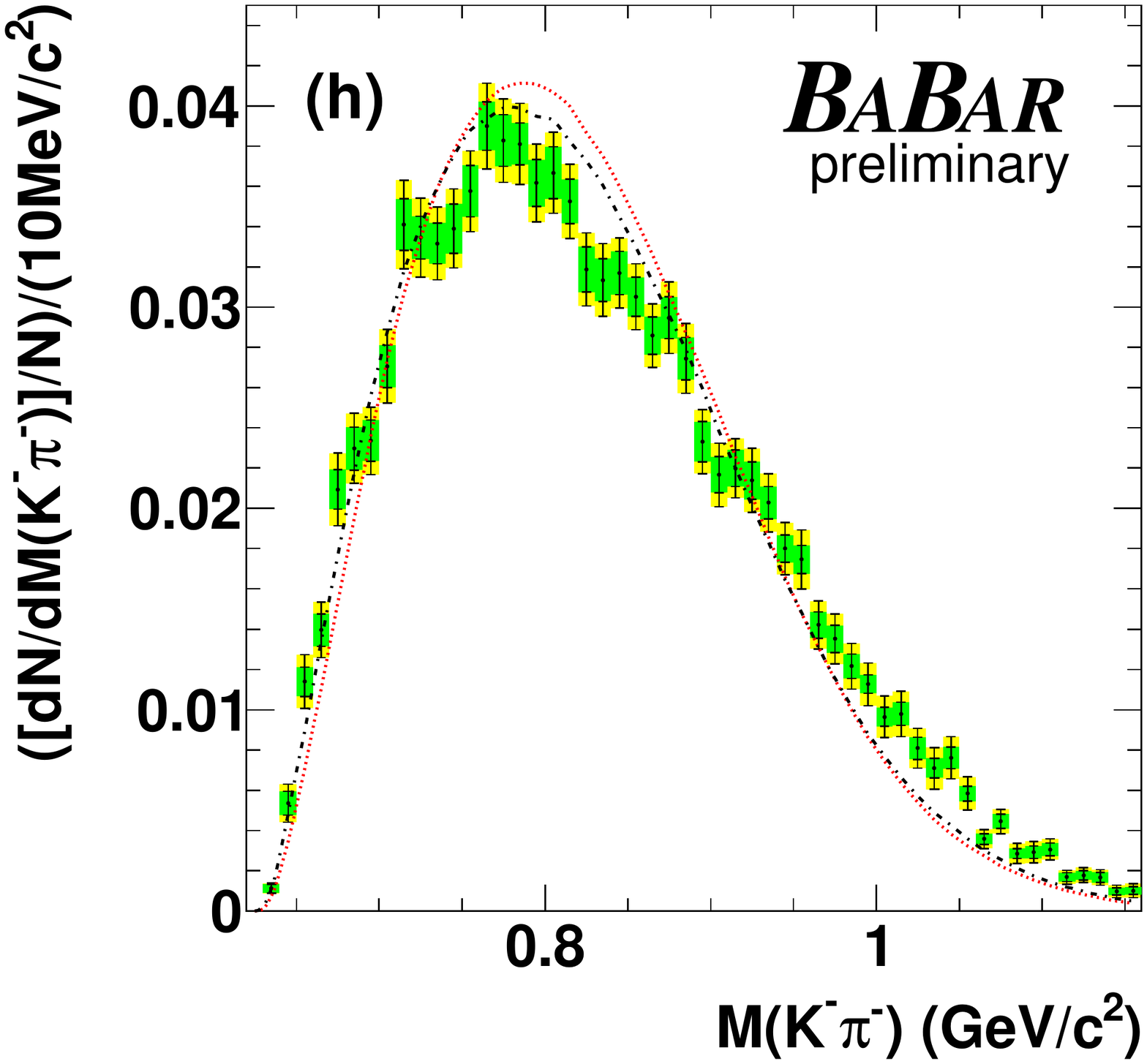}
}
\caption{ The reconstructed and unfolded invariant mass spectra for
the \tautokpk channels.  The reconstructed invariant mass
distributions for (a)
$M(K^{-}\pi^{-}K^{+})$, (b)  $M(K^{-}K^{+})$, (c)
$M(\pi^{-}K^{+})$ and (d) $M(K^{-}\pi^{-})$ are
presented in the first row. In the second row, the (e)
$M(K^{-}\pi^{-}K^{+})$, (f)  $M(K^{-}K^{+})$, (g)
$M(\pi^{-}K^{+})$ and (h) $M(K^{-}\pi^{-})$  unfolded invariant mass
spectra are shown.
For the reconstructed mass plots, the data is represented by the
points with the error bars representing the statistical
uncertainty. The blue (dark)  histogram represents the non-$\tau$
background MC, the green (medium dark) histogram represents the $\tau$
backgrounds excluding the \tautohhh cross-feed  which
are represented by the yellow (light)  histogram. For the unfolded
mass plots,
the  data is represented by the
points with the inner error bars (green) representing the statistical
uncertainty and the outer error bars (yellow)
representing the statistical and systematic uncertainties added in
quadrature. The integral of the unfolded distribution has been
normalized to 1. The black dashed line is the generator level MC
distribution used in the \babar\ simulation. The red dotted line is
the
CLEO tune for \tauola 2.8~\cite{Golonka:2003xt}.}
\label{figure1_kpik}
\end{center}
\end{figure*}

\begin{figure*}
\begin{center}
\resizebox{300pt}{99pt}{
\includegraphics[bb = 0pt 0pt 570pt 535pt]{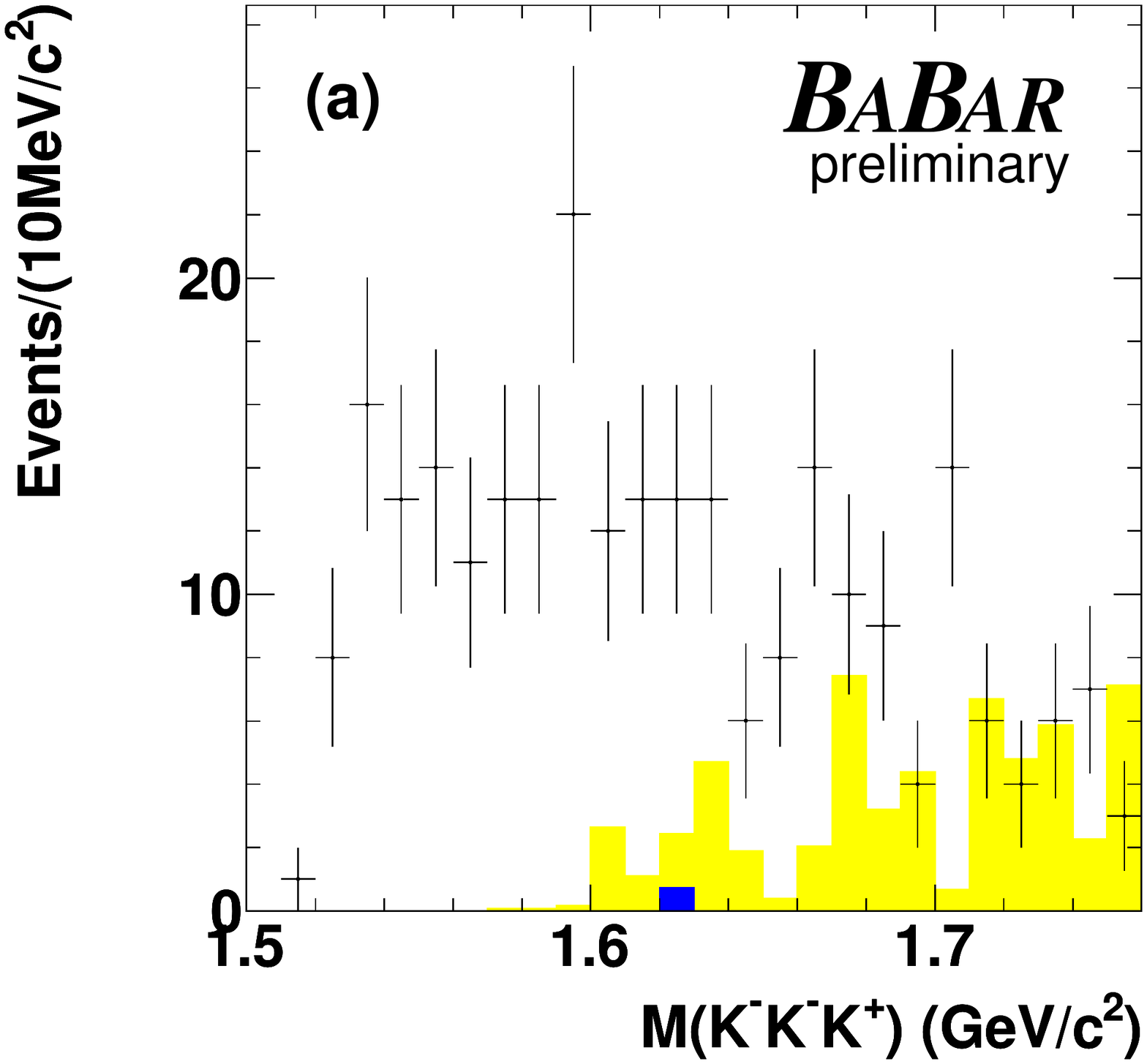}
\includegraphics[bb = 0pt 0pt 570pt 535pt]{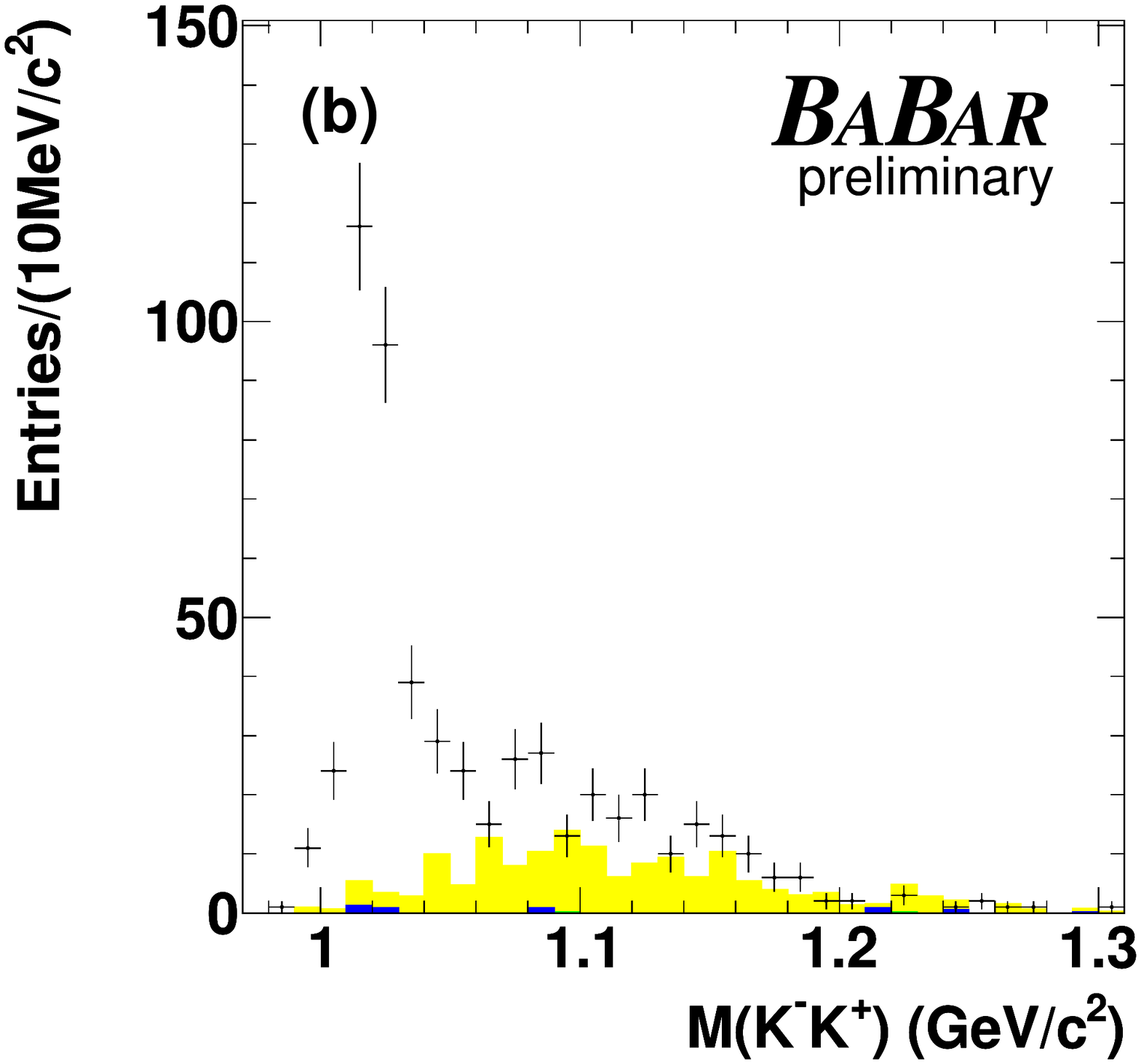}
\includegraphics[bb = 0pt 0pt 570pt 535pt]{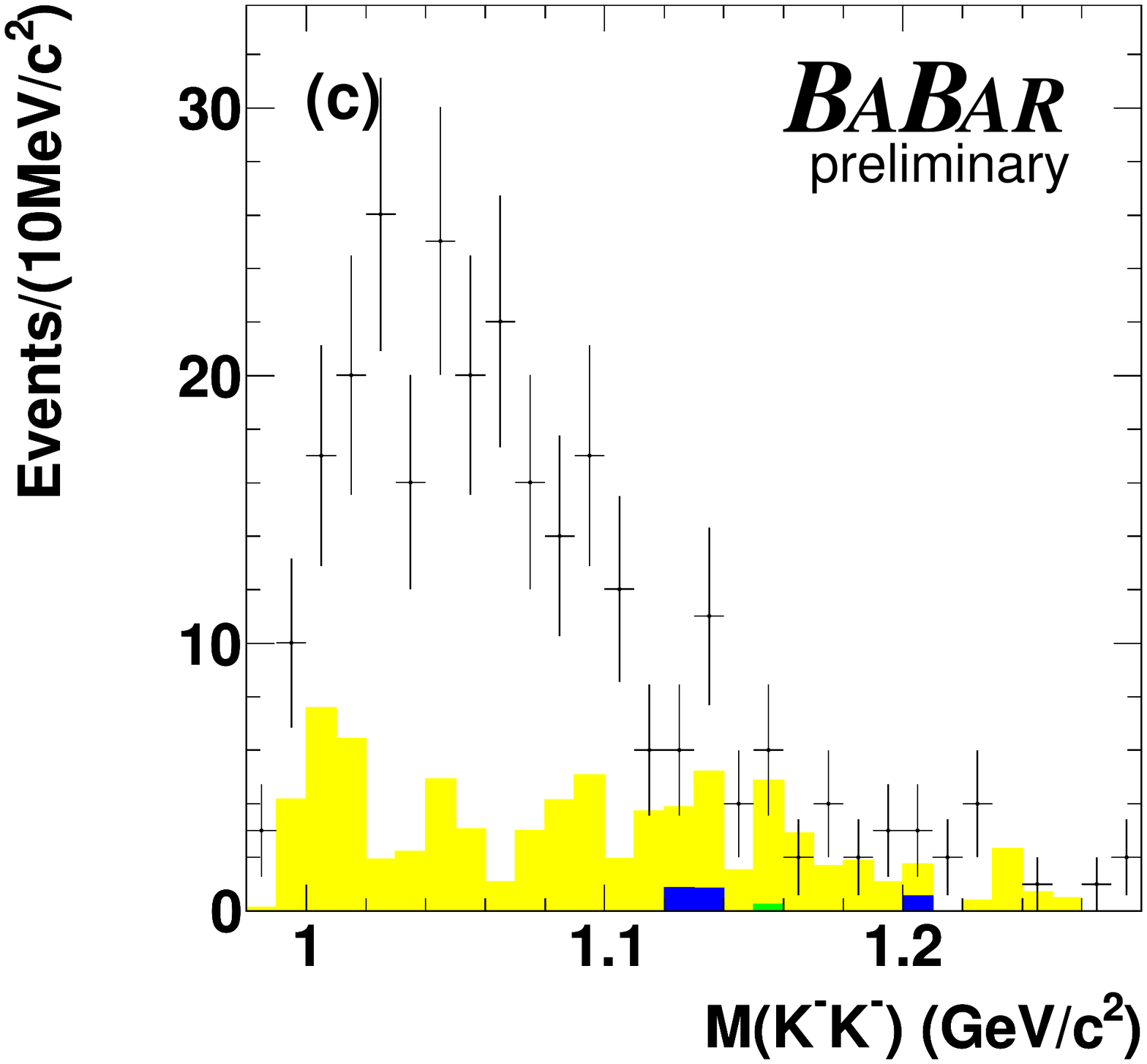}
}
\resizebox{300pt}{99pt}{
\includegraphics[bb = 0pt 0pt 570pt 535pt]{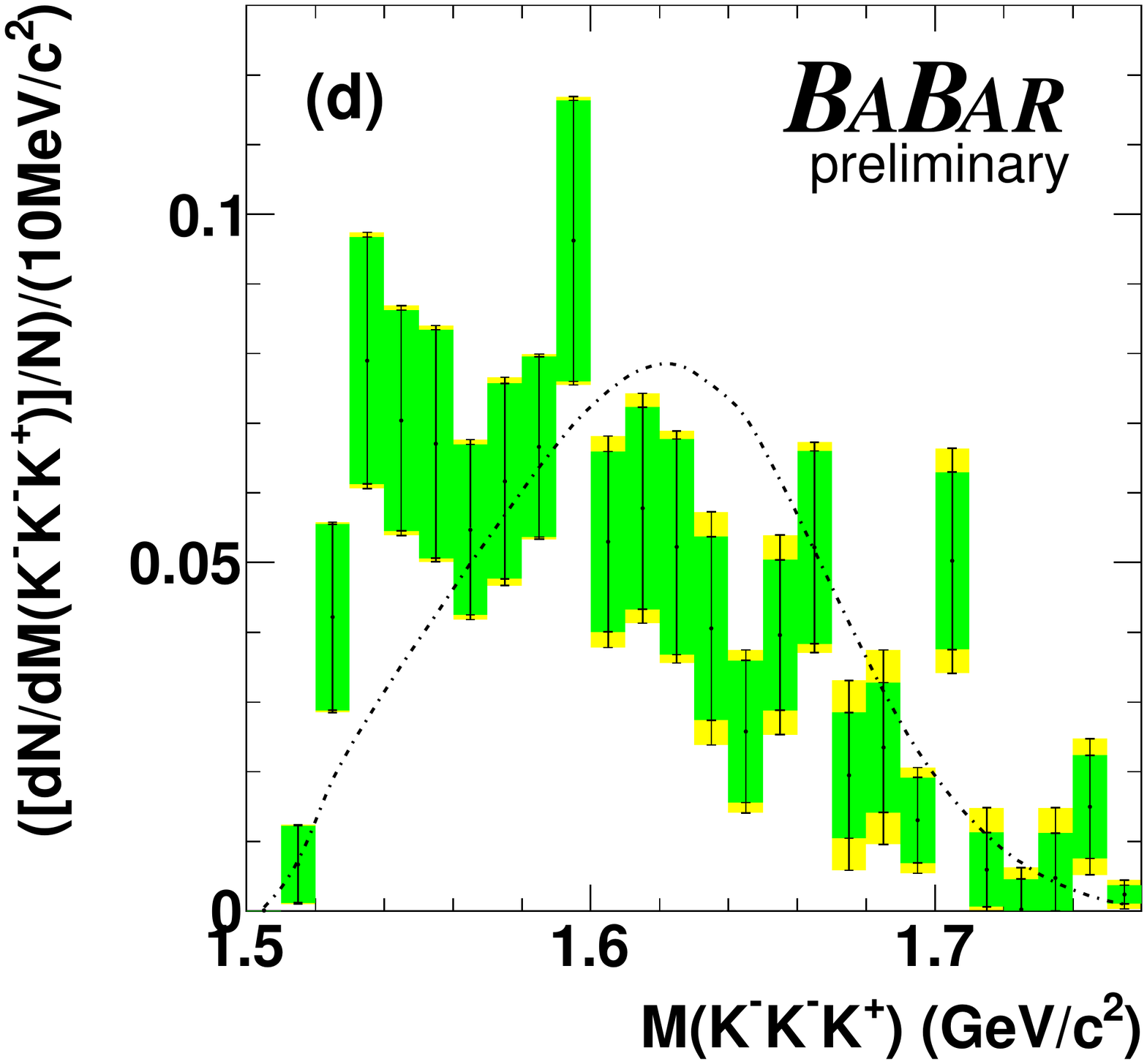}
\includegraphics[bb = 0pt 0pt 570pt 535pt]{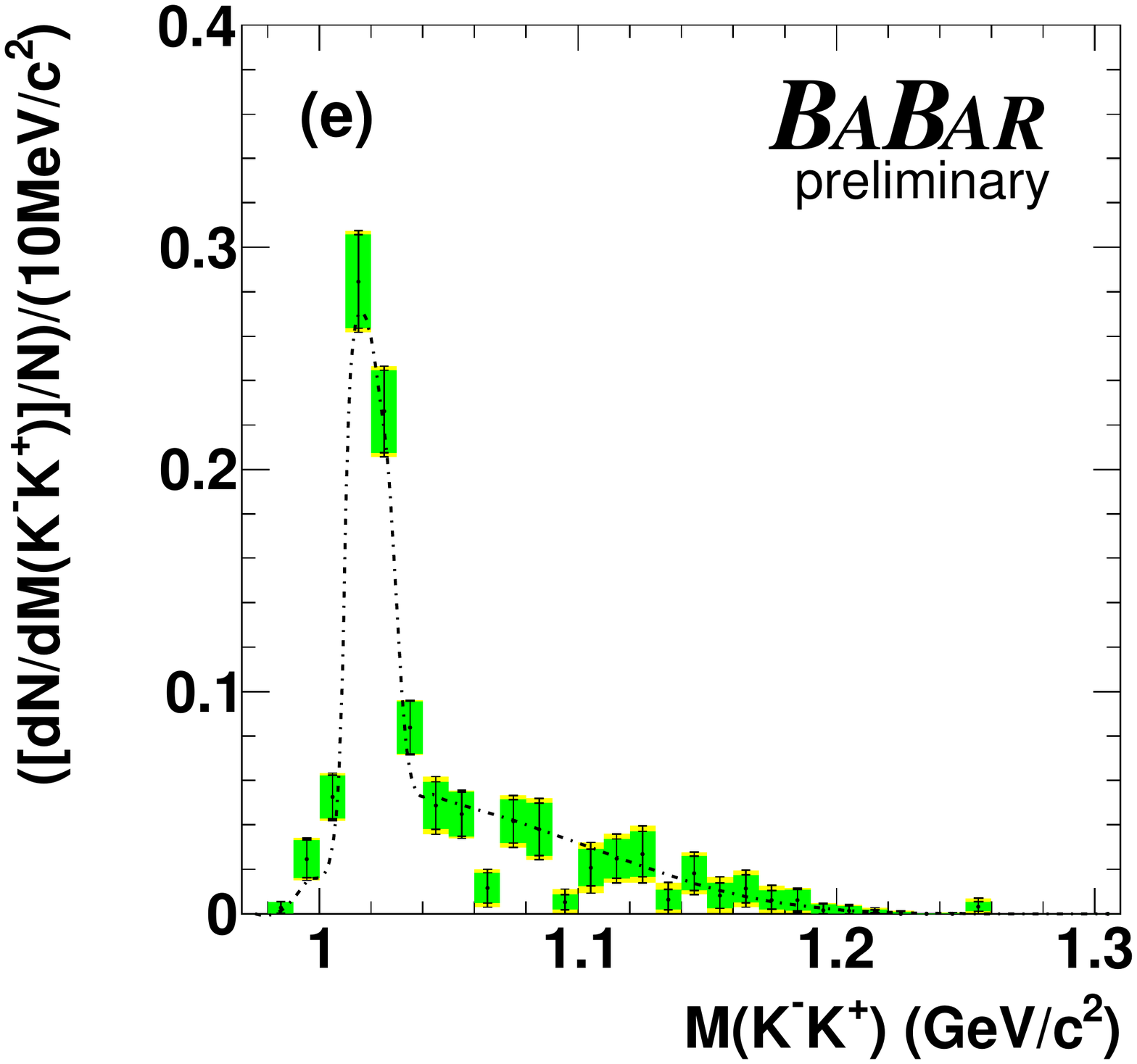}
\includegraphics[bb = 0pt 0pt 570pt 535pt]{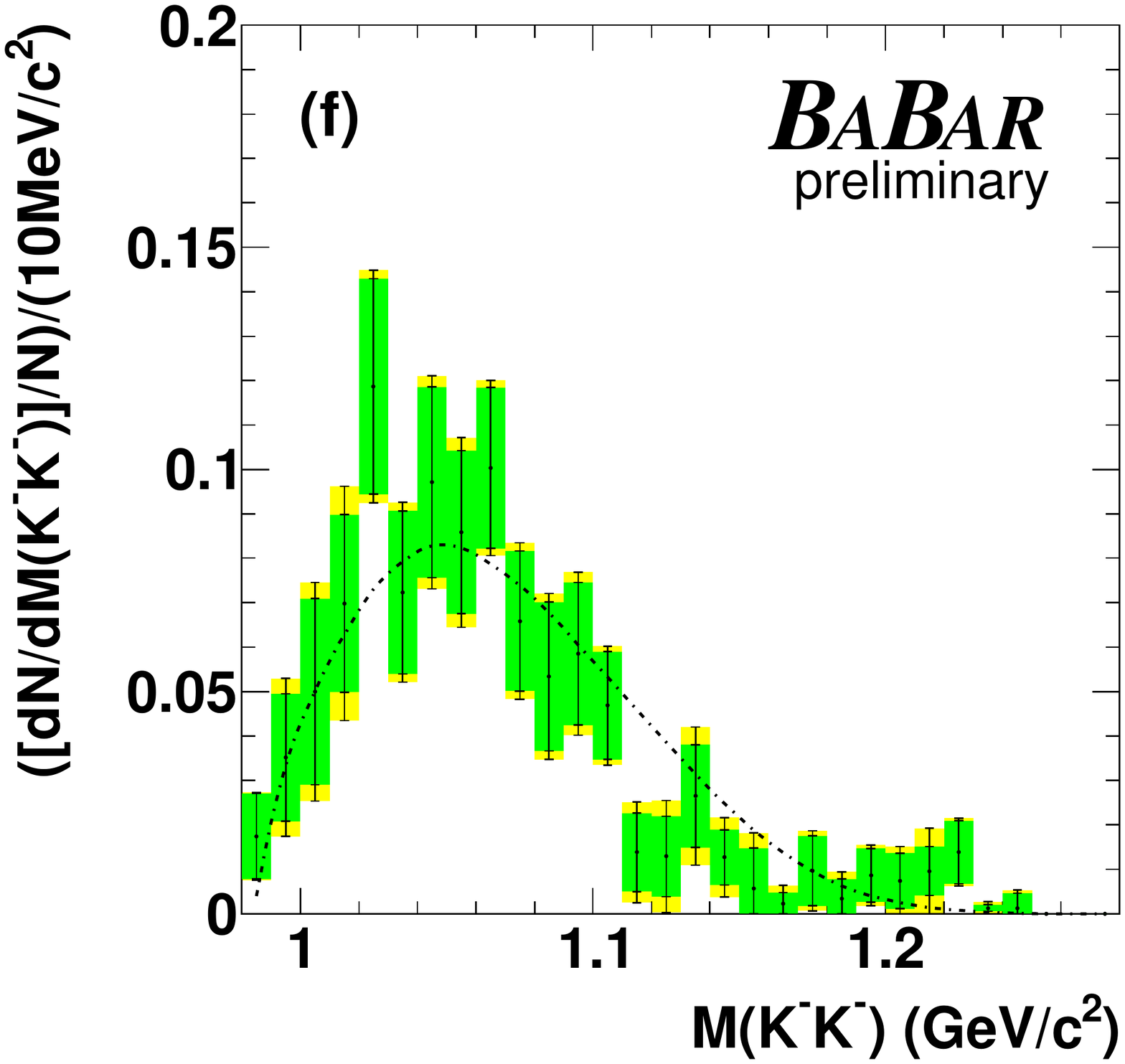}
}
\caption{ The reconstructed and unfolded invariant mass spectra for
the \tautokkk channels.  The reconstructed invariant mass
distributions for (a)
$M(K^{-}K^{-}K^{+})$, (b)  $M(K^{-}K^{+})$ and (c)
$M(K^{-}K^{-})$ are
presented in the first row. In the second row, the (d)
$M(K^{-}K^{-}K^{+})$, (e)  $M(K^{-}K^{+})$ and (f)
$M(K^{-}K^{-})$ unfolded invariant mass spectra are shown.
For the reconstructed mass plots, the data is represented by the
points with the error bars representing the statistical
uncertainty. The blue (dark)  histogram represents the non-$\tau$
background MC, the green (medium dark) histogram represents the $\tau$
backgrounds excluding the \tautohhh cross-feed which
are represented by the yellow (light)  histogram. For the unfolded
mass plots,
the  data is represented by the
points with the inner error bars (green) representing the statistical
uncertainty and the outer error bars (yellow)
representing the statistical and systematic uncertainties added in
quadrature. The integral of the unfolded distribution has been
normalized to 1. The black dashed line is the generator level MC
distribution used in the \babar\ simulation which assumes that
\tautokkk decays entirely through \tautophik.}
\label{figure1_3k}
\end{center}
\end{figure*}

The measured and unfolded invariant mass distributions for the
\tautoppp,
\tautokpp, \tautokpk and \tautokkk can be seen in Figures
\ref{figure1_3pi} to \ref{figure1_3k}. Due to the good resolution of
the \babar\ detector, the unfolding procedure has a minor impact on
the
invariant mass distributions. This is in contrast to the subtraction
of the backgrounds which have the most significant impact on these
distribtuions.
The main backgrounds come from the
 cross-feed from the other \tautohhh decays in which
 a pion or kaon has been mis-identified and from  events with an extra
 $\pi^0$. The cross-feed backgrounds are estimated to be
(0.85$\pm$0.01) for the \tautoppp channel, (38.5$\pm$0.2) for the
 \tautokpp channel, 
(2.9$\pm$0.1) for the \tautokpk channel and
(27.7$\pm$3.0) for the \tautokkk channel, where the uncertainties are
 from MC statistics.
 The background fractions from  events with an extra
 $\pi^0$ in the candidate samples
are estimated to be (3.6 $\pm$ 0.3)\% from $\tau^{-} \rightarrow
\pi^{-}\pi^{-}\pi^{+}\pi^{0}\nu$ in \tautoppp, (2.3 $\pm$ 0.4)\%
 from $\tau^{-} \rightarrow K^{-}\pi^{-}\pi^{+}\pi^{0}\nu$ in
\tautokpp, (0.4 $\pm$ 0.1)\% from $\tau^{-} \rightarrow
 K^{-}\pi^{-}K^{+}\pi^{0}\nu$
 in \tautokpk  and less than 5.0\% from $\tau^{-}
 \rightarrow K^{-}K^{-}K^{+}\pi^{0}\nu$ in \tautokkk.
The non-$\tau$ backgrounds comprise less than 0.5\% of the events for
 each channel. 

\section{Discussion and Conclusion}
\label{DiscandCon}

In this paper, we have presented unfolded invariant mass spectra using
Baysian unfolding for the decay modes \tautoppp, \tautokpp, \tautokpk 
and \tautokkk.
Additional studies have been conducted that confirm the \babar\ branching
fraction measurement \cite{ourpaper} using control samples to
cross-check all the main backgrounds. The \tautokpppz and \tautokpkpz are
also used to validate that the $K/\pi$ mis-identification rates are
consistent within the assigned systematic uncertainties. 
These invariant mass distribtuions are essential for determining the strange and
non-strange spectral density function and for the extraction of
$|V_{us}|$. Work is ongoing with T. Przedzinski, P. Roig,
O. Shekhovtsova, Z. Was to improve the modeling of these decay modes
for the LHC and for the next
generation B-Factories using this data.

\section{Acknowledgements}
We are grateful for the 
extraordinary contributions of our \pep2\ colleagues in
achieving the excellent luminosity and machine conditions
that have made this work possible.
The success of this project also relies critically on the 
expertise and dedication of the computing organizations that 
support \babar.
The collaborating institutions wish to thank 
SLAC for its support and the kind hospitality extended to them. 
This work is supported by the
US Department of Energy
and National Science Foundation, the
Natural Sciences and Engineering Research Council (Canada),
the Commissariat \`a l'Energie Atomique and
Institut National de Physique Nucl\'eaire et de Physique des Particules
(France), the
Bundesministerium f\"ur Bildung und Forschung and
Deutsche Forschungsgemeinschaft
(Germany), the
Istituto Nazionale di Fisica Nucleare (Italy),
the Foundation for Fundamental Research on Matter (The Netherlands),
the Research Council of Norway, the
Ministry of Education and Science of the Russian Federation, 
Ministerio de Ciencia e Innovaci\'on (Spain), and the
Science and Technology Facilities Council (United Kingdom).
Individuals have received support from 
the Marie-Curie IEF program (European Union) and the A. P. Sloan Foundation (USA).

Additional support for I. M. Nugent was provided by the Alexander von Humboldt
Foundation.

\nocite{*}
\bibliographystyle{elsarticle-num}
\bibliography{paper}

\end{document}